\documentclass{esub2acm}

\newcommand{\isDRAFT}{0}
\newcommand{\doAppendix}{1}
\newcommand{\doTOC}{1}

\newcommand{\appendixtexxSCCSID}{appendix.texx 3.2}
\newcommand{\cookiebibSCCSID}{cookie.bib 3.4}
\newcommand{\cookietexSCCSID}{cookie.tex 3.11}

\usepackage{ifthen,alltt,url}

\ifthenelse
    {\isDRAFT = 0}
    {%
	\newcommand{\Draft}[2]{#1}
    }
    {%
	\newcommand{\Draft}[2]{
	    \ifthenelse{#2 = 0}
		{#1 --- \textbf{DRAFT} \today}
		{\textbf{DRAFT} \today --- #1}
	}
    }

\newcommand{\ifAppendix}[2] {%
    \ifthenelse%
	{\doAppendix > 0}%
	{%
	#1%
	}%
	{%
	#2%
	}%
}

\begin{document}

\ifthenelse
    {\doTOC > 0}
    {%
	\tableofcontents
    }
    {%
    }

\title{HTTP Cookies:  Standards, Privacy, and Politics}

\author{David M. Kristol\\
	Bell Labs, Lucent Technologies}

\begin{abstract}
How did we get from a world where cookies were something
you ate and where ``non-techies'' were unaware of ``Netscape cookies''
to a world where cookies are a hot-button
privacy issue for many computer users?
This paper will describe how HTTP ``cookies'' work,
and how Netscape's original specification evolved into
an IETF Proposed Standard.
I will also offer a personal perspective on how
what began as a straightforward technical specification
turned into a political flashpoint when it tried to
address non-technical issues such as privacy.
\end{abstract}




\ifAppendix
   {\def\permission{\par Copyright \copyright~2001, Lucent Technologies.
   	All Rights Reserved. \\
	\today
	    ~\cookietexSCCSID
	    ~\appendixtexxSCCSID
	    ~\cookiebibSCCSID
	}
   }
   {}

\begin{bottomstuff}
\begin{authinfo}
\address{600 Mountain Ave., Murray Hill, NJ  07974}
\end{authinfo}
\permission
\end{bottomstuff}
\markboth{\protect\Draft{David M. Kristol}{0}}
	{\protect\Draft{HTTP Cookies:  Standards, Privacy, and Politics}{1}}
\maketitle

\section{Introduction}
The topic of HTTP ``cookies'' has become at least slightly familiar
to many Internet users.
Articles in the popular press now regularly mention cookies in conjunction
with privacy concerns.
However, cookies were in use for over two years before they first achieved
notoriety, and some of that notoriety emerged around the same time as
the appearance of the first formal standard for cookies,
which had previously been informally described on Netscape's
Communications Corporation's web site.

The cookie standardization process began in April, 1995, with
a discussion on \cite{www-talk}.
In December of that year, the IETF undertook to write
a cookie standard.
After a series of Internet Drafts got published in connection
with extensive public discussion on \cite{http-wg}
(and after noticeable delays due to IETF process),
RFC 2109~\cite{rfc2109}, \emph{HTTP State Management Mechanism}, was published in February, 1997
as an IETF Proposed Standard.
Technical and political concerns immediately emerged that
led to further discussions and revisions (and delays) and that
finally culminated, in October, 2000, in the publishing
of RFC 2965~\cite{rfc2965}.

In this paper I describe what cookies are, how they work, and
how applications use them.
I also
\ifAppendix
    {relate}
    {briefly relate\footnote{
    See \cite{cookie-tm} for an extended history.
    }
    }
the history of how cookies came to be standardized,
and how the protracted process of standardization interacted with other
forces in the explosive early evolution of the World Wide Web.
We participants in the standardization process unexpectedly
found ourselves at the intersection of technology and public
policy when the proposed standard raised concerns about privacy.
As co-editor of the specification,
I'll reflect on what happened.

\section{What are Cookies?  Why are they Useful?}

Any discussion of cookies must begin by answering two
questions:\footnote{
Inevitably there's a third question:  Where did the term
``cookie'' come from, anyway?
\textit{Magic cookie}, or just \textit{cookie}, is a computer jargon
term with a long and honorable history; it refers to an
opaque identifier that gets passed back and forth between
different pieces of software~\cite{hackdict}.
}
What are they?
Why are they needed?
The answers require a modest understanding of how the World Wide Web
(WWW or ``Web'') works, which the next section provides.

\subsection{An Introduction to Hypertext Transfer Protocol (HTTP)}
The Hypertext Transfer Protocol (HTTP~\cite{rfc2616})
provides the foundation for the Web,
and cookies are an addition to HTTP.
When a user clicks on a (hypertext) link in a web browser,
the browser (sometimes referred to as ``client'' or
``user agent'')
typically connects to the web server identified by the Uniform
Resource Locator (URL) embedded in the link and sends it a request
message, to which
the server sends a response message.
Then,
after receiving the response,
the browser disconnects from the server.
Because the client makes a new connection for each request,
the server treats each request as though it were the first one
it had received from that client.
We therefore consider the request to be ``stateless:''
each request is treated completely independently of any previous
one.\footnote{
Newer clients and servers are able to maintain a connection
for more than one request-response cycle, but the essential
stateless behavior remains.}

Statelessness makes it easier to build web browsers and servers,
but it makes some web {applications} harder to write.
For example, it would have been much harder to create the now-ubiquitous web
shopping applications if they could not keep track of what's in
your shopping basket.

HTTP requests (responses) comprise three parts:
\begin{enumerate}
  \item{a request (response) line;}
  \item{request (response) headers, which provide meta-information; and}
  \item{the request (response) entity itself.}
\end{enumerate}

The header meta-information provides both control information
for HTTP and information about the entity being transferred.
Information about cookies gets conveyed in such headers.

Here is an example request.  The first line is the request
line.  The remaining lines are request headers.
There is no entity for a \verb,GET, request.
\begin{alltt}
GET / HTTP/1.1
Accept: image/gif, image/x-xbitmap, image/jpeg,
  image/pjpeg, application/vnd.ms-powerpoint,
  application/vnd.ms-excel, application/msword, */*
Accept-Language: en-us
Accept-Encoding: gzip, deflate
User-Agent: Mozilla/4.0 (compatible; MSIE 5.5; Windows 98;
   Win 9x 4.90)
Host: aleatory.research.bell-labs.com:80
Connection: Keep-Alive
\textnormal{---blank line---}
\end{alltt}

The corresponding response might look like this.
\begin{alltt}
HTTP/1.1 200 OK
Date: Thu, 25 Jan 2001 16:40:54 GMT
Server: Apache/1.3.12 (Unix)
Last-Modified: Fri, 05 Jan 2001 23:38:49 GMT
ETag: "121be7-15d-3a565b09"
Accept-Ranges: bytes
Content-Length: 1706
Content-Type: text/html
\textnormal{---blank line---}
\textnormal{---HTML entity here---}
\end{alltt}

Note that the request and response information is readable text,
although the entity transferred need not be.
``Web pages'' are often encoded in Hypertext Markup Language (HTML),
as in this example (although the actual HTML content
has been omitted).

\subsection{How Do Cookies Work?}

Web-based applications often use cookies to maintain state in the otherwise
stateless HTTP protocol.
As part of its response,
a server may send arbitrary information, the ``cookie,''
in a \texttt{Set-Cookie} response header.
This arbitrary information could be anything:  a user identifier,
a database key, whatever the server needs so it can continue
where it left off.
Under normal circumstances (and simplifying greatly), a cooperating
client returns the cookie information verbatim
in a \texttt{Cookie} header, one of its request headers,
each time it makes a new request to the same server.
The server may choose to include a new cookie with its responses,
which would supersede the old one.
Thus there is an implied ``contract'' between a server and client:
the server relies on the client to save the server's state and
to return it on the next visit.

To correct a frequent misstatement in early press stories, cookies
do not arise from some insidious invasion of your computer
or hard drive by an external intruder.
Rather, your browser stores only those cookies it receives
from a server it has visited.
(However, it will be clear later that your browser
may visit servers on your behalf without your knowing
it and store cookies from them on your computer.)

A ``cookie,'' then, is the piece of information that the server
and client pass back and forth.
The amount of information is usually small, and its content
is at the discretion of the server.
In general simply examining a cookie's value will not
reveal what the cookie is for or what the value represents.

Restricting the client to return a cookie just to the server from which it
was received is very limiting.
Organizations often have multiple servers, and those servers need
to have access to the same state information so that, in the
aggregate, they can provide a service. 
Therefore, when it sends a cookie to a client,
a server may specify, in a constrained way, the set of
other servers to which a client may also send the cookie
in subsequent requests.

The server can tell the browser to return cookies only in
requests to specific parts of its site, as identified
by the URLs.
Provided applications use different parts of the ``URL space''
(for example,
\url{http://shop.com/application1} and \url{http://shop.com/application2})
servers (or sets of servers) may thus host multiple cookie-based
applications.\footnote{
Suggestive domain names (such as
\texttt{shop.com}) in the examples
merely convey the role the names play in the example.
They do not imply that anything described in such examples applies to
any site that might exist with that domain name.
}

\subsection{Proxies}

Users can configure their browsers to use
an HTTP \emph{proxy} for a variety of reasons, such as to improve performance,
or because their Internet Service Provider (ISP)
 or company requires them to do so.
An HTTP proxy is an intermediary that accepts a request from a client
and forwards it to a server, then receives the resulting response
and forwards it back to the client.
A typical proxy accepts connections from multiple clients and can
connect to any server.
A pure HTTP proxy does not save the results of the requests and responses
it forwards and poses no problems for applications that use cookies.

A \emph{caching proxy}, however, may store responses.
The purpose of a caching proxy is to reduce network traffic and
response latency:  the caching proxy may be able to return
the same response to client 2 that it previously returned to client 1 for the
same request, without the need to forward the second request to
the origin server.
However, in some cases returning the same response to
both clients is the wrong thing to do, such as when the response
for client 1 contains personalized content, or when the response
is time-dependent.
In such a case, the origin server must send response headers
that direct any proxies between the server and the client (there may
be zero or more proxies) not to cache the content, or to
revalidate the response with the origin server before returning
the response to the client.

Cookies and caching proxies can interact in undesirable ways, and
cookie-based web applications must take that possibility into account.
For example, a shopping site may allow a page with product information
to be cached, but it should not allow a \verb,Set-Cookie, response header
(and its associated cookie) to be stored with it.
On the other hand, the shopping site should suppress the caching
of any pages that contain personal information, such as shipping
information or the contents of a shopping basket.

\subsection{Why Cookies?}

Cookies make it easier to build stateful web applications,
but they are not essential to achieve that purpose.
To accomplish much the same thing,
a server can, for example, embed state information in URLs,
use hidden fields in HTML forms,
or use the client's Internet Protocol (IP) address.
But these approaches are failure-prone.
As Section \ref{ip-addr-not-good} describes,
IP addresses are an unreliable way to identify a user or computer.
If URLs or forms are used,
the state information is not part of the protocol; rather it is
contained within the user-accessible information that the server
returns to the user.
If a user clicks on a Back button in the browser, the user's state
would roll back to what it had been for the earlier page.
For a shopping application, this behavior would have the effect of
removing items from the shopping basket.
Moreover, both approaches lend themselves to mischief:  a user
can easily capture the text of the URL or form fields, edit it, and resubmit
the information to the server, with unpredictable results.
Finally, embedding state information in URLs is very unfriendly to
caches, and web caches are considered valuable for reducing
network traffic and, thereby, congestion.

\subsection{How Are Cookies Used?}

Cookies have a variety of uses, some of which are controversial.
I've already described how they can make it possible
(actually, strictly speaking, they make it \emph{easier})
to implement shopping applications.

Cookies can also be used to store ``login'' information for sites
that provide personalized access, so you don't have to keep entering
your name and password each time you visit.\footnote{
\cite{rfc2964} deprecates using cookies to store such
information under some circumstances.}

A web site can also use cookies to track which pages you visit on the site.
The site's administrators may want to use the cookies
to better understand how users navigate the site.
With such an understanding, they can organize the site so
the most popular information is in places that are easier for users to find.

Ordinarily a web site (\texttt{example.com})
cannot distinguish a particular user over time;
at best it can tell what IP address your computer has.
\label{ip-addr-not-good}
However, that IP address often does not identify you uniquely:
\begin{itemize}
\item If you use an HTTP proxy, the web site will ``see'' the proxy's
IP address, not your computer's.  Thus all of the proxy's users will
appear to be one user to a server.
\item If you use an ISP that provides you
with a temporary IP address each time you connect to it, your IP
address could be different when you visit \texttt{example.com} at different
times, and you would appear to the server as different users.
\end{itemize}

A cookie that's stored on your host computer is indifferent to the
path by which your computer connects to \texttt{example.com}.
The ``cookie contract'' stipulates that your computer return
its cookie to \texttt{example.com} when you visit it again,
regardless of what your IP address is (or which ISP you use).
On a single-user computer like a PC, the cookie thus identifies
the collection of all users of the computer.
On a multi-user computer, the cookie identifies the user(s) of a
particular account.

\textit{Identifies} does not necessarily mean that
\texttt{example.com} somehow knows your name, address, or other personal information.
Unless you explicitly provide personal information, all that
\texttt{example.com} can do is assemble a list of URLs on its site
that you (or, rather, the user of a particular computer or
account) have visited, as identified by a cookie.
Of course, if you \emph{do} supply personal information to \texttt{example.com},
perhaps to register for some service or to order merchandise, that
information can be associated with the URLs you visited.
As I'll discuss later, this ability to monitor your browsing
habits and possibly to associate what you've looked at with who
you are is at the heart of the privacy concerns that cookies raise.

\section{The IETF Standards Process}

Because the cookie specification and the HTTP specification
have emerged from the Internet Engineering Task Force (IETF) standards process, it's essential
to understand the IETF's hierarchical
organization and how its standards process works.

The IETF evolved out of the
earliest days of the Arpanet to become the \textit{de facto}
Internet standards body.
At the lowest level are ``members.''
Unlike most other standards bodies, however, where the body is
an internationally sanctioned group and participants are
national representatives, IETF is an open organization
whose members comprise literally anyone
who is willing to participate constructively in IETF activities.
There are no membership cards or dues.
While they are often employed by corporations with a
keen interest in the ultimate shape that IETF standards take,
members are expected to develop standards on their \emph{technical merits}
alone, and by custom they are assumed to speak for themselves, and
not their employers.\footnote{
That's not to say that there aren't differing opinions about
``technical merits,'' which may reflect corporate interests.
}

Members typically participate in one or more \emph{working groups} of
interest, such as the HTTP Working Group.
A working group (WG) organizes itself,
chooses one or more chairpeople,
and draws up a charter, which
identifies work items and a schedule by which they will be completed.
Working groups are expected to have a limited lifetime,
on the order of 2-3 years, although the work items in the charter
usually span 12-18 months.
Most of the work of a working group gets done on email mailing lists,
which are open to anyone to join, and which must have a public archive.
Occasionally small numbers of working group members meet face-to-face
to work out particularly knotty issues.
Otherwise, the only time most members actually \emph{see} one another,
if ever, is
at IETF meetings, held three times a year.
One or more attendees take meeting notes, which then get published
on the IETF's web site after the meeting.
The availability of meeting notes, open mailing lists, and list archives
helps keep the process transparent to anyone interested.

The standards that the IETF produces begin life as an
\emph{Internet Draft} (I-D).
Anyone may submit an I-D to the Internet Drafts Administrator of the IETF.
Typical
I-Ds are part of a working group's business, and some of
those are \textit{standards track}, as opposed to, say, \textit{informational}.
I-Ds usually have a six-month expiration:
if no action is taken on an I-D, it vanishes.
Two typical actions are for
the author to submit a revised I-D to supersede the first, or for
the IESG (see below) to approve an I-D to advance
along the standards track.

Attendees at working group sessions at the meeting are
expected to have read the applicable current Internet Drafts.
The IETF sets a cut-off date for submitting
I-Ds in advance of each meeting.
Therefore, a flood of new I-Ds typically gets announced by the IETF
just before a meeting, as authors try to make their latest
version available.
The cut-off ensures an adequate amount of time to review the I-Ds.

Groups of Working groups comprise \emph{Areas}.
For example, the HTTP Working Group was part of the Applications Area.
Each Area has one or two \emph{Area Directors}, who are experienced IETF
members.
The Area Directors as a group comprise the
\textit{Internet Engineering Steering Group} (IESG).
IESG members administer IETF process, and
they monitor the activities of the working
groups and, among other things,
watch for similar work in different areas that perhaps should
be coordinated.

The \emph{Internet Architecture Board} (IAB) comprises senior members
of the IETF who guide the overall evolution of Internet standards
and adjudicate disputes about IESG actions.

A typical IETF standard's life-cycle begins with an initial I-D.
The I-D undergoes vigorous scrutiny and discussion by a working
group on its mailing list, which results in a cycle of revised I-Ds
and further discussion.
When the working group reaches ``rough consensus,''
the chair issues a \textit{Last Call} for working group comments.
Assuming all of those are addressed adequately by the author(s),
the chair recommends that the IESG consider the I-D 
to be a \emph{Proposed Standard}.
The IESG then issues its own, IETF-wide, Last Call for comments.
If there are comments, the author(s) will revise the I-D
and restart the discussion cycle, although at this point
there usually are but few comments.
Once the IESG approves the I-D to be a Proposed Standard,
 it gets submitted to the
\emph{RFC Editor}, who edits and formats the document and assigns
it an RFC (Request for Comments) number.
Once published as an RFC, a document never changes.
It can only be superseded.
Indeed, as a specification progresses through the IETF process,
a newer RFC often supersedes a previous one.

The IETF emphasizes ``rough consensus and running code.''
Note that ``consensus'' does not mean ``unanimity.''
The process requires that all voices must be heard, but not all
voices must be heeded.

Interested parties are expected to implement a 
specification once it becomes a Proposed Standard.\footnote{
Indeed, they often begin to do so before then,
and their implementation experience often provides feedback
for the evolution of the specification even before it becomes an RFC.
}
Before the RFC can progress to the next stage of the
process, \emph{Draft Standard},
there must be evidence that at least two
independently developed interoperating implementations exist.
This requirement ensures, first, that the specification
is clear enough that two or more implementors, working
independently, interpret the specification the same way.
Second, the requirement demonstrates that the pieces so-developed
can actually communicate with each other,
which, of course, is essential for any useful networking
protocol!
Finally, after further review and implementation experience,
a mature specification advances to
\emph{Full Standard} (or just \emph{Standard}).

\section{RFCs 2109 and 2965:  A Brief History}

\subsection{In the Beginning\ldots}

When the World Wide Web first made its way into the public's consciousness
in 1993, the web browser of choice was \emph{Mosaic}, from
The University of Illinois's National Center for Supercomputing Applications
(NCSA).
Mosaic offered no support for a state mechanism.
Early applications that wanted or needed to support state had to
find an unsatisfactory workaround, possibly among those mentioned
earlier.

The first publicly available version of the Netscape Navigator
browser (September, 1994)
supported state management~\cite{lou-1}, although that fact
was not well known at the time.
The mechanism had been introduced at the request of one of Netscape's
customers to provide the kind of stateful mechanism we now recognize.
Lou Montulli at Netscape wrote the original specification, and he chose
the term ``cookie.''
Cookies solved the problems identified earlier:
applications worked correctly even in the face of a user's
page navigation; and
cookies were part of the protocol, not part of the content, and thus they were
less accessible to a user.
Applications that used cookies were more robust than those using alternatives.

It now seems hard to imagine the Internet ``landscape'' in April, 1995,
when the cookie story begins.
Corporate and government web sites had started to proliferate.
The technical community actively discussed the possibilities
of what we now call ``e-commerce.''
ISPs had started to appear and to offer software bundles
that incorporated Mosaic or Navigator.
Some early adopters had Internet access at home.\footnote{
America Online (AOL), with two million subscribers,
had yet to offer direct Internet access, and did not until October, 1997.
AOL now has over 29 million subscribers.
}
More people had Internet access at work, but even email access
was relatively uncommon outside the technical community.
Amazon.com would not open for business for three months.
No one had used the phrase ``dot-com.''

The current cookie standard reflects the interplay of technical issues,
personalities, IETF procedures,
corporate influences, and external political influences.
Table \ref{timeline} summarizes the important events
in the standardization timeline.
Although the process took a long time, note that significant
``dead time'' occurred following Last Calls.
\ifAppendix%
    {The Appendix}
    {\cite{cookie-tm}}
recounts in
considerable detail the evolution of the cookie specification
through two RFC cycles.
In the sections immediately below
I give a condensed version that highlights
the key issues.

\begin{table}[h]
\centering
\begin{tabular}{|ll|} \hline
September, 1994	&	Netscape Navigator 0.9 beta, includes cookie support	\\
April, 1995	&	discussions begin on \cite{www-talk}		\\
August, 1995	&	\texttt{State-Info} I-D				\\
December, 1995	&	state management sub-group of HTTP WG forms	\\
February, 1996	&	\texttt{state-mgmt-00}, first public cookie I-D	\\
June, 1996	&	working group Last Call on \texttt{state-mgmt-02}	\\
August, 1996	&	IESG Last Call on \texttt{state-mgmt-03}	\\
February, 1997	&	RFC 2109, \textit{HTTP State Management Mechanism}	\\
March, 1998	&	working group Last Call on \texttt{state-man-mec-08}	\\
June, 1999	&	IESG Last Call on \texttt{state-man-mec-10}	\\
April, 2000	&	new IESG Last Call on \texttt{state-man-mec-12}	\\
October, 2000	&	RFC 2965, \textit{HTTP State Management Mechanism}	\\
		&	RFC 2964, \textit{Use of HTTP State Management}	\\
\hline
\end{tabular}
\caption{Timeline of HTTP State Management Standardization}
\label{timeline}
\end{table}

\subsection{RFC 2109:  December, 1995 to February, 1997}

By late 1995, three proposals for adding state to HTTP were
circulating in the technical community.
Because the HTTP Working Group was more concerned with producing an
HTTP/1.1 specification to solve urgent needs,
Larry Masinter, as chair of the group, asked
the parties interested in state management to form a sub-group to recommend
a single approach to the rest of the WG.
As author of one such proposal, I agreed to head up the
``state sub-group,'' and a group of eight people,
including Lou Montulli, the author of Netscape's
specification~\cite{NS}, began to meet by email and
conference call.
After considering the alternatives, we soon decided to adopt
Netscape's underlying mechanism, while preparing a more
precise specification.

\cite{NS} provides rules whereby a cookie can be shared among
multiple servers, based on their domain names.
We identified two problems that this
cookie sharing mechanism could enable\ifAppendix%
    {\footnote{See Section \ref{domain-attr} for more details.}}%
    {}%
:
1) Cookies can ``leak'' to servers other than those intended by
the originating server.
2) A server in a domain can cause a denial-of-service attack,
either inadvertently or intentionally,
by sending cookies that will disrupt an application that runs
on another server in the same domain (``cookie spoofing'').
We worded the specification to try to minimize how widely
cookies could proliferate, subject to the (implicit) constraint that
the control be based on domain names.

In February, 1996, we identified what we felt was a considerable
threat to privacy, \emph{third-party cookies}, or ``unverifiable
transactions.''\ifAppendix%
    {~\footnote{More detail can be found in Section \ref{3rd-party}.}}
    {}
A transaction, or request, is ``verifiable'' when the user can
tell beforehand where it will go.
A browser can receive third-party cookies if it loads a page from
one web site, loads images (such as ads) from another web site,
and the latter web site sends a cookie with the image.
Our concern was that, whereas a user could well expect a cookie
from the first web site, she has no reason to expect, or even
to know, that her browser will visit another web site
(through an unverifiable transaction) and receive
a cookie from it.
We added wording to the specification that either outright prohibited
a browser from accepting third-party
cookies (``cookies in unverifiable transactions''),
or that permitted a browser to accept them, provided
they were controlled by a user-controlled option whose default value
was to reject them.

By late April, 1996, the sub-group had prepared an I-D for review
by the entire WG.
After some revisions, there was a WG Last Call in June, some small
revisions, and an IESG Last Call in early August.
In October, the IESG expressed concern that, in essence, suggested
the specification was too lenient with respect to identifying
which transactions were ``verifiable.''
Keith Moore, an Applications Area Director, and I worked out
compromise wording with the IESG to convey the idea that
the inspection mechanism described in the specification
was at best minimally acceptable.
In December, with this change, plus another, minor one, the IESG approved
the specification to be published as an RFC, which it
was in February, 1997, as
RFC 2109,~\cite{rfc2109} ``HTTP State Management.''

Some of the threads common to the evolution of the specification
had already manifested themselves:
\begin{itemize}
\item{We were concerned about how cookies'
domain names affected their ability to proliferate
beyond their intended (or desired) targets.}
\item{We had noted a potential privacy threat in ``third-party cookies.''}
\item{The IESG pushed for even stricter language regarding
``third-party cookies'' than the WG felt was feasible, given
constraints of compatibility and what could reasonably be
demanded of an implementation.}
\item{At a particularly volatile time in the evolution of Web technology,
IETF process roughly doubled the time between
the specification's being accepted by the WG and the time it
appeared as an RFC.}
\end{itemize}

\subsection{RFC 2965:  February, 1997 to October, 2000}

RFC 2109 attempted to extend \cite{NS} while using the same
HTTP headers.
The hope was that already-deployed clients and servers could
be upgraded incrementally to use the new specification.
However, around the time that the IESG approved RFC 2109, but
before it got published, a compatibility issue surfaced.
We found that Netscape Navigator and Microsoft Internet Explorer
(MSIE)
behaved differently in the face of the attributes we had
introduced as part of RFC 2109.
Clearly the WG would have to revise the specification.

\subsubsection{Fixing the Incompatibility}

Because the two extant major browsers disagreed on how to
treat unfamiliar attributes, we were inexorably led to
introduce one or more new headers to resolve the problem.
We discussed several different approaches, all of which
entailed putting the ``new'' attributes in a new header,
where they would not confuse the code that handles existing
headers.\ifAppendix%
    {\footnote{ More complete discussion is in Section \ref{incompat}. }}%
    {}%

\subsubsection{Unverifiable Transactions and Certified Cookies}

The publication of RFC 2109 resulted in articles about cookies
appearing in the trade and popular
press and to heated protests from the Web advertising
networks that emerged while the RFC was being written and discussed.
Because many of the networks had developed business models
that relied on third-party cookies to do targeted advertising,
they felt the RFC's mandate to disable third-party cookies
by default was a threat to their business.
However, the WG generally supported the RFC's default,
noting that the RFC's restrictions on third-party cookies
would affect the advertisers' \emph{business models} that
relied on tracking users, not the advertising business itself.

These discussions about third-party cookies
led to a proposal of ``certified cookies.''
A certified cookie would assert how the web
server would use the cookie, and it would be signed cryptographically
by an auditing agency.
A user could configure her browser to specify what kinds of
uses of cookies she is comfortable with, and the browser
would automatically accept or reject cookies, whether they
were from third parties or not, based on
the configuration.
The WG's goal was to layer the certified cookies mechanism
on top of the regular cookie mechanism to enhance the
(default) third-party cookie ban and other cookie controls.

\subsubsection{Deadlock and Resolution}

Through early 1997, the WG attempted to resolve
the issues described above (and others\ifAppendix%
    {; see Section \ref{history-2}}%
    {}%
).
By August, however, discussions had become circular, repeating
earlier arguments and mingling technical and social (privacy)
issues.
We were making no progress.

As a way out of the impasse, we embarked on a two-part strategy
to proceed.
I would remove, temporarily, the parts of the specification
concerning ``unverifiable transactions,'' letting us focus
on the purely technical part.
Once we agreed on the technical part, we would re-introduce the political part
and try to reach further consensus, at which point we should be done.

By February, 1998, we had achieved consensus on the technical part
of the specification.
When I subsequently added back the ``unverifiable transactions''
language, surprisingly there were no further comments, and Working Group
and IESG Last Calls quickly followed.
However the resulting specification, with minor modifications,
languished for two years.
Apparently
the IESG felt the
need to set stronger guidelines for the use of cookies
than the prospective (new) RFC contained.\footnote{
The new RFC did not differ materially from RFC 2109
in its privacy provisions.
However, the composition of the IESG had changed
in the intervening 3 1/2 years.}.
Only when this set of guidelines (\cite{rfc2964})
was written and accepted could the cookie specification
be published as RFC 2965.

\section{Privacy/Politics}

The cookie specification may have been the first IETF standard
at the intersection of technology and privacy to get widespread
public notice, some of which I'll describe below.
As the Internet moved from research plaything to public plaything
to vital communications infrastructure,
the IESG began to expect all RFCs to include a
thoughtful ``Security Considerations'' section.\footnote{
\cite{rfc1543},
superseded by \cite{rfc2223}, which
came \emph{after} RFC 2109, called for a ``Security Considerations''
section, but
many RFCs said there were no security issues.
}
Privacy was considered an element of security in this context.
Indeed, the longest delays incurred during the standardization
process of the two cookie RFCs were due to the tension between
the IESG, which pushed for \emph{even stricter} privacy safeguards than
those two RFCs contained, and the HTTP Working Group, which
could achieve even rough consensus only with slightly weaker safeguards.

\subsection{Federal Trade Commission}
The U.S. Federal Trade Commission (FTC) convened a workshop
on consumer privacy in June, 1996.
Among the topics discussed was the possible use of
the World Wide Web Consortium's (W3C)
Platform for Internet Content Selection (PICS)~\cite{PICS}
``to facilitate automatic disclosure of privacy policies
and the availability of consumer choice regarding
the use of personal information.''\cite{FTC96}

In March, 1997, just weeks after RFC 2109 appeared,
the FTC announced another
Consumer Information Privacy Workshop to be held in June~\cite{FTC97}.
Among the topics discussed were consumer online privacy,
industry self-regulation, and technology that could be used
to enhance privacy. 
In a comment letter to the FTC regarding the workshop,
Peter F. Hartley, Netscape's Global Public Policy Counsel
wrote:
\begin{quotation}
2.14 Interactive technology has evolved since June 1996 to address
many of the privacy concerns expressed regarding cookies and what
information was placed on a user's computer and with what notice
and consent. Software manufacturers and open technical standards
bodies have produced innovations that enable users to have more
control over cookies and how web site operators are able to place
information on one's computer. At this point in time many web site
operators and related third parties are reviewing the technical
standards concerning cookies. These improvements and changes in
cookie technology will be implemented in upcoming versions of
Netscape products. However, as Netscape is an open standards
company we cannot at this time specifically detail the latest version
of this cookie standard until the report of the most recent IETF
meeting is released and reviewed.~\cite{hartley} 
\end{quotation}
Clearly cookies and their privacy implications
had become an issue of federal public policy.\footnote{
Indeed, in June, 2000, the Clinton administration
banned cookies from federal Web sites unless their was
a ``compelling need.''~\cite{nytimes-010417}\,
Three bills before the U.S. Congress as of April, 2001,
refer to ``cookies.''
}
Since the FTC workshop, cookies have become a frequent topic
in the popular press.

\subsection{W3C and P3P}

As an outgrowth of the 1996 FTC meeting, members of
the W3C began to discuss
a PICS-like mechanism for privacy preferences~\cite{cranor-email}.
In May, 1997, W3C
formed the Platform for Privacy Preferences Project (P3P).
According to its information page, P3P
\begin{quotation}
is emerging as an industry standard
providing a simple, automated way for users to gain more control over
the use of personal information on Web sites they visit. At its most
basic level, P3P is a standardized set of multiple-choice questions,
covering all the major aspects of a Web site's privacy policies.~\cite{p3p-goal}
\end{quotation}

P3P defines a mechanism whereby a web site can send its privacy
policies in a well-defined form and using a well-defined vocabulary.
The machine-readable form of the information facilitates
automatic processing, making
it possible to compare it to a user's privacy preferences.
P3P would appear to provide the mechanism needed to support certified
cookies.

\subsection{Industry Self-Regulation}
In response to the FTC's hearings,
the U.S. government and the advertising industry
entered a dialog in which
the industry created a self-regulating mechanism\footnote{
The industry formed the Network Advertising Initiative
\url{http://www.networkadvertising.org}.
}
to try to avert threatened government regulation of web
sites' privacy policies
and drew up a set of principles.
Separately, TRUSTe formed in 1996 
to create a Privacy Seal that would attest to an organization's
privacy practices.
Clearly the issue of public trust in web sites's privacy practices
has emerged as an important issue, whether regulated by the
government or the industry~\cite{governing-trust}.

\subsection{Third-Party Cookies}
The advertising networks protested that the cookie standard
threatened their business.
In truth, what the standard really threatened, by disabling
third-party cookies by default, was a business
\emph{model}.
The Web advertising business comprises two parts:
\emph{deciding} what ad to return in response to a request
and actually sending it.
Unconstrained third-party cookies allow the \emph{decision} to be targeted
more precisely to a user.
But eliminating third-party cookies would not prevent an
advertising network from returning an ad.

Targeted ads are a symptom of something more troubling, profiling, and
the accumulation of profiles has been at the heart of the controversy.
On one side are the advertising networks, who maintain that
the assembling of a user's (anonymous) profile makes it possible
for them to present ads that are more likely to interest the user.\footnote{
Of course, they can also charge more money for them, or they
can hope for more revenue through a higher ``click-through rate,''
or they can hope to sell the profiles, perhaps after linking
them with personally identifying information.
}
On the other side are privacy advocates (and users),
who question whether advertising networks are entitled to assemble
such profiles,
and who say
that at the very least the advertisers
must get users' permission before they do so.\footnote{
Although they express concern about profiling,
users do seem to accept the concept
of advertising as a means to support web sites.
}

The initial reaction from the advertisers to the default
setting for third-party cookies was clearly negative.
They didn't see the need to ask for permission before
setting cookies, and they felt
that asking for it would be too burdensome anyway.
As the weight of public unease and the threat of governmental  
regulation grew, they showed support for techniques like
certified cookies, which matches a user's comfort level with an
advertiser's declared use, to bypass the crude all-or-nothing
enabling of third-party cookies in browsers.

It should be noted that
disabling third-party cookies does not eliminate the profiling of
users, only profiling done by third parties, and even then
web bugs~\cite{web-bugs} can be used for much the same purpose.
Web sites that serve their own ads can still create profiles
of visitors to their sites, using cookies within their own domain.
However, the profiles are less comprehensive, because they derive
from just the one site or a set of related sites,
and users are less likely to be surprised by such profiling,
because they consciously visited the site.

The advertisers frequently insisted that they only compiled
anonymous profiles to be used for targeting their ads,
and that they had no plans to match the profiles with
personally identifiable information.
However,
\begin{quotation}
[T]he merger of database marketer Abacus Direct with online ad
company DoubleClick hit front pages and sparked a federal
investigation in January 2000 when it was revealed that the
company had compiled profiles of 100,000 online users ---
without their knowledge --- and intended to sell
them~\cite{privfound-2000}.
\end{quotation}
This behavior confirmed users' worst fears and justified our
concern about third-party cookies in the standards process.

\subsection{Do Users Care?}
Do users care whether a web site tracks them?
The question has no easy answer.
One study~\cite{websidestory}
showed that users rejected fewer than 1\% of cookies
in over a billion page views.\footnote{
Privacy policies routinely go unread, too.}
This simple number may have many explanations, among them that
users:
\begin{itemize}

\item{don't know about cookies.}
\item{know about cookies but don't know how cookies
might be used to track them.}
\item{know how cookies can track them, but are unconcerned.}
\item{have inadequate means to select which cookies
they will accept, so they just give up and accept all of them.}
\item{assume that the entities collecting the information will
protect it and use it discretely.}
\item{assume governmental regulations will prevent
web sites from misusing information about them that they might collect.}

\end{itemize}

Most Web browsers enable cookies by default,
and most users won't even know they're in use.
Yet I think most users would object to the idea that a web site
uses cookies to watch what they do and where they go on the site.
I think most users would be even more uneasy that some unseen entity
uses cookies to watch all of their Web surfing activities~\cite{kafka}.
However, cookies are used by most shopping sites, a popular
activity on the Web, so it's impractical for users to disable
cookies altogether.
Moreover, the heavy use of cookies on some sites renders
them virtually unusable if a user enables cookie warnings.
Lacking easy to use selective control of cookies, 
or much motivation\footnote{
People willingly divulge information about themselves
every day through credit card purchases,
supermarket discount cards, and electronic
toll collection systems.},
the average user is most likely just to leave the default
settings alone.
Indeed, that's why advertisers who rely on third-party
cookies objected to their being disabled by default.

\section{Lessons Learned}

\begin{quote}
The best of all teachers, experience.  \emph{Pliny the Younger}
\end{quote}

I've learned some things the hard way
that may help guide other standards writers.

\subsection{Writing Standards Is Hard}
A technical specification is a contract between consenting applications.
It is amazingly hard to write a standard that says neither more
nor less than you intend, that leaves room for implementation
flexibility without resulting in incompatibility, and that is
precise enough to avoid ambiguous (mis-)readings.
The process of developing standards and letting
all voices be heard is also messy, like most democratic processes.
It gives me a new appreciation for how hard it must be to
write legislation, even ignoring the distorting influences
of lobbyists.

\subsection{Zombie Topics}
During the standardization process,
some topics never died.
Throughout the process, issues that we thought we
had resolved got raised again, and consensus had to be achieved
again.
No doubt this revivification arose as new participants entered
the discussions belatedly, and the fact that the discussions
took time to resolve highlights how rough, in fact, the consensus was.
Perhaps a Frequently Asked Questions document that summarizes
resolved issues could short-circuit unproductive rehashes.

\subsection{Reconciling Incompatible Goals}

\subsubsection{Domain-matching Rules}

We began work on the cookie specification intending to be
as compatible as possible with \cite{NS} while being more precise and
while trying to control inadvertent or malicious cookie sharing.
The domain matching rules reflected this intent, but the intent
was ultimately impossible to achieve.
\ifAppendix
    {As described in Section \ref{domain-attr}, the}
    {The}
domain matching rules implicitly assume
properties of the domain name system that do not actually exist,
and it is therefore impossible to allow cookies to be
shared as widely as desired but not \emph{too} widely.
Perhaps the certified cookie specification will solve this
problem better.

\subsubsection{Compatibility and Deployment}

HTTP is readily extensible if new features derive from
adding headers, as long as clients and servers that ignore\footnote{
Which is what they are supposed to do if they see
an unfamiliar header.
}
the
headers will function correctly.
Some thought must also be given to what an
older proxy would do with the header.

Changing the behavior of existing headers to extend
an \emph{existing feature} is harder, as we discovered.
When the state sub-group began work, only Netscape Navigator
supported cookies, and we could base our decisions on
how Navigator did things to arrive at a compatible solution.
By the time RFC 2109 was complete, Microsoft Internet
Explorer was widely deployed, and, as luck would have it, Navigator
and MSIE were incompatible in an important way.

A solution
that uses new headers to solve the incompatibility problem
poses deployment
problems, which is why the state sub-group had avoided
introducing new cookie headers in the first place.
Ultimately the new-header solution should
supersede the old, but meanwhile there is a transition
period during which both headers must be used.
The duplicate headers impose extra costs on servers that
send them, yet the servers experience little benefit
from sending them.
Web site operators that conscientiously choose to support the
new specification could well find that few client programs
support it, and thus they
would incur the added cost for a long transition period.
With little incentive for servers to support the new specification,
there would be correspondingly little incentive for browser vendors
to add support for it.

\subsection{Speak Now, or Forever Hold Your Peace}
A pattern emerged in the cookie specification process:
After considerable discussion, I would produce a presumed final I-D.
At Last Call, a sudden flurry of
substantive comments would appear that should have been made earlier.
Apparently the Last Call provided the stimulus that was necessary to
rouse people to take a serious
look at the specification.

Sometimes it's hard to identify all the constituencies that
need to participate.
The advertising networks felt they had been blind-sided by IETF
process, and that the resulting standard was a threat to their
business.
The problem was that the advertising networks
were unaware the IETF
was producing a cookie standard
that might affect their business, even though the process was open.
Perhaps we were remiss not to inform them.
However, we were largely unaware of their activities, at least initially.
In any case, the (``rough consensus'' of the) working group and the IESG
steadfastly supported
the ``third-party cookie'' default that was their
principle objection.

\subsection{How Wide is the World Wide Web?}
We chose the default setting for third-party cookies because
we felt it served the privacy expectations of users,
especially European users, who, we inferred from
European Union recommendations, might have high expectations.
In 1999, a European Union Working Party Recommendation
stated its concern
\begin{quotation}
about all kinds of processing operations which are presently being performed by software
and hardware on the Internet without the
knowledge of the person concerned and hence are ``invisible''
to him/her
\end{quotation}
and went on to mention specifically
``the cookies mechanism as currently implemented in the common browsers.''~\cite{eu:wp17:1/99}

Surely, we reasoned, vendors would choose to take such
concerns into account for \emph{all} users.
Evidently we reasoned wrong.
Vendors have steadfastly supported the advertising
industry, leaving third-party cookies enabled by default.

\subsection{Technical Decisions May Have Social Consequences}
When I began working on the cookie specification, I thought
I was trying to write a technical specification, and, frankly,
I hoped my \verb,State-Info, proposal would prevail.
However, lurking within both that proposal and \cite{NS}
were social policy issues I had not anticipated.
The simpler \verb,State-Info, proposal specifically 
mentioned privacy; \cite{NS} implicitly considered
privacy through its domain-matching rules.
The final specification clearly addressed social issues.

Of course it's not uncommon for technology to create or enable
social consequences, and technologists are often ill-equipped,
and perhaps are even inappropriate parties,
to deal with the fallout.
Consider just one recent example.
Cell phone use has led to the desire to provide emergency (``911'')
service, just as is provided for wired phones.
To provide such service, cell phone providers need to know
reasonably accurately the location of the cell phone user.
However, users of such phones may thereby lose the anonymity
of movement they would have had without such phones.
Moreover, businesses stand ready to try to sell the cell
phone users their products when they are nearby.
The technology began with a worthy
intent, but other, less noble, uses may evolve.

The task of reconciling social and technical choices can
be hard even when the social component is acknowledged
from the beginning~\cite{cranor-reigle}.
Looking back, it's clear that the social ramifications of
the cookie specification took on more importance, and were harder
to resolve, than the purely technical ones.
Or, looking at it another way, we discovered that our
apparently technical decisions had social consequences.

Cookies are inherently neither good nor bad.
They can enhance web applications, and they can be used
to invade privacy;
technology alone cannot distinguish good uses from bad.
In fact, just labeling a use as ``bad'' is highly
subjective.
Even members of the state management sub-group had
mixed opinions. 

Who gets to decide these issues?
Sometimes technologists do, just by the capabilities they build
into, or leave out of, their work.
Sometimes technologists throw up their hands and say, ``Not my
problem.''
But society increasingly holds
businesses accountable for
the secondary effects of their products,
and technologists can't simply ignore the effects
of what they build.
Ultimately I think few of the conflicts at the
intersection of technology and society can be resolved
by wholly technological means.
Because the resolution must balance the needs and desires of
various constituencies, the conflict must in the end be
resolved through the political process.

\subsection{How to Do it Better}
In this section I reflect on how things might have been done differently.

\subsubsection{Involve the stakeholders}
Successful standards involve the stakeholders.
Although the IETF nominally comprises \emph{individuals}, the reality
is that those individuals work for companies, and those companies
have a stake in the standardization process.
Thus it was prudent for us to involve
Netscape, particularly, where cookies originated,
and Microsoft.
Indeed, when we began our work, Netscape's representative
did participate.
But Microsoft was not yet a factor in Web software
and did not participate early on.

Evidently the specification veered in a direction Netscape could
not support,
although we frequently sought feedback about Netscape's plans
and buy-in for the evolving specification.
A representative from Microsoft belatedly joined the discussions
around the time RFC 2109 appeared, but he was often unhappy
with the details of both RFC 2109 and its successor, particularly
the domain-matching rules and third-party cookie default.

Web site operators are another major stakeholder.
They are unlikely to adopt changes that cost them time and money
unless they gain something in return.
Although many sites probably rely on the software they buy
from vendors to ``do the right thing,''
they should be increasingly concerned by possible
federal legislation concerning privacy.
Although none of the privacy legislation has yet become law,
there is clearly growing sentiment to ``do something,''
because the voluntary compliance with self-regulation has been dismal.
Web sites should probably expect that they will be \emph{required}
to notify users how they plan to use personal information,
and they may be required to let
users ``opt-in'' rather than ``opt-out.''\cite{S.2606-106th}
The \verb,CommentURL,
feature of RFC 2965\ifAppendix
    {~\footnote{See Section \ref{commenturl}.}}%
    {}%
,
coupled with well-designed
user interfaces in common browsers, would probably satisfy the
requirements of opt-in information collection.
However, the browser vendors have shown little enthusiasm for the feature.

\subsubsection{Separate policy and mechanism}
It's common to argue that mechanism should be separated from policy,
and that the policy rules should
be specified in a separate document from the one that
specifies the mechanism.
Indeed, I have described how we were unable to solve, through technology
alone, the privacy problems that cookies might cause.

The specification we developed included many intrusions
of \emph{policy} into the \emph{mechanism}.
For example, the domain matching rules dictate which cookies to accept
and which to reject.
However, the political reality was that the IESG would probably
not have accepted a specification that side-stepped the
implications of a mechanism unburdened by privacy considerations.

\subsubsection{Avoid lily-gilding}

We fell prey to the temptation to add features that \emph{seemed}
worthwhile without actually getting agreement that they would
actually be used.
(On the other hand, IETF process would require that unimplemented features
be removed from the specification before proceeding to
Draft Standard or beyond.)
Sometimes the discussions of these extra features, however worthy,
distracted from the larger goal of making a usable specification
available quickly.

\section{Conclusions}

\subsection{Timing Matters}
\begin{quote}
Delay is the deadliest form of denial.  \emph{C. Northcote Parkinson}

Not to decide is to decide.  \emph{Anonymous}
\end{quote}

When the state management sub-group began its work in December, 1995,
Netscape's Navigator browser dominated
the market, Microsoft's Internet Explorer
barely existed, e-commerce was nascent, and
advertising networks scarcely existed.\footnote{
DoubleClick incorporated in January, 1996 and first came up
in our discussions in June.
}
We attempted to standardize something that closely
resembled Netscape's under-specified cookies, but we felt
the need to mitigate the privacy threats that we perceived
could be mounted by cookie applications.
And we felt a sense of urgency to produce a specification
quickly, to keep pace with the evolving HTTP/1.1 specification
and, by producing a tight specification,
to provide a level playing field for all browser vendors.

Unfortunately, events did not unfold as we might have hoped.
Although what ultimately became RFC 2109 was essentially complete
seven months after we began, for a variety of reasons
the RFC itself did not appear for yet another seven months.
During those fourteen months, MSIE emerged as a serious
competitor to Navigator, e-commerce began to blossom, and advertising
networks had become common in the Web ecosystem.

Each of these factors made the world different from
when we began our work.
MSIE became much more than a marginal browser, and we could not
ignore the incompatibilities that we had discovered.
E-commerce applications were becoming sophisticated, and their software
investment reduced the likelihood that they would be willing
to switch to IETF cookies.\footnote{
On the other hand, to the extent that support for IETF cookies,
those following RFC 2109 or 2965,
found its way into shrink-wrapped e-commerce  applications, deployment
could conceivably occur quickly.
}
And the advertising networks that had business models that depended on
creating profiles by using third-party cookies
felt that the language in RFC 2109
was a dagger aimed at the heart of their business.\footnote{
Although RFC 2109 said third-party cookies were generally forbidden.
a browser could have an option to enable them, provided
the default setting for such an option was ``off.''
}

Of all of these issues, the one that indirectly got the most attention
was the third-party cookie default.
The attention was indirect in that, shortly after RFC 2109 was
published, a flurry of media articles appeared about cookies
and whether they were a threat to privacy.
I was occasionally asked at the time whether I in fact knew of any privacy
violations that had actually taken place, and I had to answer, ``No,''
that they were, as far as I knew, hypothetical.
Then again, I didn't expect someone doing those things to admit it.
Events
since then,
such as the proposed merger of DoubleClick and Abacus Direct databases,
have shown that the threats were anything but hypothetical.

The browser vendors showed, through their actions, that they were
unwilling to change the third-party default to ``off.''
In my opinion, that choice was hardly surprising.
At a time when Microsoft and Netscape were giving away browsers
to try to achieve market dominance, while at the same time
they were selling servers, both vendors were most likely to
heed their paying customers, not the people who
got programs for free.
And the people with the money \emph{wanted} advertising,
they \emph{wanted} to use advertising networks,
and most advertising networks \emph{wanted} to be able to do targeted
advertising.
Targeted advertising was easiest to do using third-party cookies,
and the server/browser vendors were unlikely to anger their paying customers
by disabling third-party cookies.

The $3~1/2$ year delay between RFC 2109 and RFC 2965 would appear to
have rendered the latter RFC moot.
While I was personally committed to seeing the specification through to its
conclusion, I think the Web has evolved too much for web sites
to revise their applications to use it and for vendors to change
the clients and servers to use it.
Moreover, the end of the ``browser wars'' means
there's much slower turnover of browsers, and it would take a
very long time for RFC 2965-compliant browsers to penetrate
the user community in a meaningful way.

\subsection{On the Bright Side\ldots}
Having painted such a bleak picture, I must however point
out some bright spots.
Netscape introduced modest cookie controls in their
browsers, apparently in response to the discussions leading to RFC 2109,
including the ability to disable third-party cookies,
although the default remained to allow them.
Many products, both commercial and free- or shareware, now make it
possible to control which cookies to accept and which to ignore.
Even Microsoft released patches to Internet Explorer to provide
more extensive cookie control facilities, but then they
backed off~\cite{privfound-2000}.

The issuance of RFC 2109 helped ignite a public discussion of cookies,
their uses and abuses.
That discussion has helped raise the general public's awareness
of privacy issues in general
and has made privacy a governmental policy issue.
Web sites feel public and governmental pressure to explain
their privacy policy.
When they violate that policy (toysmart.com~\cite{wired-toysmart}),
reduce its protections (Amazon.com~\cite{privfound-2000}),
or plan to join anonymous and personal data (DoubleClick and
Abacus),
they suffer in the public press.\footnote{
\cite[p.9]{dbnation} contains several even earlier examples of
public outrage at privacy intrusions.
}

Discussion of other issues continues as well.
For example, should web sites collect
personal information and then give users the opportunity not
to have the information retained (``opt-out'') or
must users be asked ahead of time (``opt-in'') before the
sites collect the information?
Is it fair for a web site to post a privacy policy
that they ``can change from time to time'' without notifying users,
and which therefore requires a user to check that policy
on a regular basis?

\subsection{Summary}
RFC 2965, \textit{HTTP State Management Mechanism}, took $5~1/2$ years
to become a Proposed Standard, and yet
the major vendors largely ignore it.
Therefore its development would, at first glance, seem to
have been a colossal waste of time.
This paper has explained why it took so long and presents
a case study of how the
collaborative IETF process works.
The fact that the standard may be largely ignored
has more to do with other factors than with its technical merit.
Moreover, the surrounding discussions of privacy considerations
may, in the long run, prove to have been more important for society
and the technical community than the technical issues.

\subsection{Why Did I Stick With It?}
Why should I continue to work on the cookie
specification for over five years
in the face of all the delays and fulminations,
particularly when my company had no stake in the outcome,
and even more particularly as the reality sank in that
the specification might well never be deployed widely?
The answer is complicated, and more than likely even \emph{I}
don't fully comprehend why.

For a start, I didn't expect the process to take so long.
I had hoped that I could wrap up my involvement when RFC 2109
appeared.
Then the incompatibility came to light, and I felt the need
to address it.
Probably equally significant, however, was the outspoken
criticism of a small number of people who seemed bent on
delaying or sabotaging the specification and the process, one of whom
more or less said to me, ``[My employer, a major vendor,] will 
never support this standard, so why are you bothering to keep
working on it?''
Feeling I was being bullied made me more determined to persist,
and I didn't like to see an attempt to bully the IETF, either.

\section{Acknowledgements}
My thanks go to Neal McBurnett, for his probing review of the paper;
to Koen Holtman, for challenging my faulty memory;
to the anonymous reviewers, who forced me to think
harder about what it was I really wanted to say;
to Balachander Krishnamurthy, for encouraging me to write the paper;
and to Avi Silberschatz, for providing more support than either he or I intended.

\bibliographystyle{esub2acm}
\ifthenelse{\isDRAFT > 0}{\smallskip \cookietexSCCSID \nocite{sccs-id}}{}

\newif\ifdoAppendix

\ifAppendix{\doAppendixtrue}{\doAppendixfalse}

\ifdoAppendix

\newpage

\appendix
\section{History of RFC 2109, HTTP State Management Mechanism}
\label{history-1}
The following sections comprise a roughly chronological
recitation of how the current cookie specification evolved.
I take this approach for several reasons:
\begin{itemize}
\item{It presents a detailed case history of how one standard evolved.}
\item{It makes it easier to explain why various sections of the
standard say what they say.}
\end{itemize}

\subsection{The State Sub-Group Forms}

Events were moving quickly in the Web technical community
in 1995
as people embraced this new, exciting medium.
The IETF HTTP Working Group (WG) had formed in October, 1994,
with a goal, among other things, to produce an
Internet Draft (and RFC) that would describe
HTTP/1.0 as it then existed.
(Developers had been using an older document by Tim Berners-Lee as the
standard reference.)

By the December, 1995, IETF meeting, the Web was recognized
as a ``success disaster'':
its explosive growth was stressing the Internet in two important
dimensions:
\begin{enumerate}
  \item{Web network traffic was growing fast,
  and HTTP's one-request-per-connection
behavior was very network-inefficient.}
  \item{As big consumer brands started to register domain names and create
web sites for those brands, they used one IP address for each such name at a time
when there was widespread fear in the IETF community that the
32-bit IP version 4 address space would soon be exhausted.}
\end{enumerate}

The IAB urged the HTTP working group to produce
a new standard quickly that would address these problems.

Meanwhile, the topic of keeping state to track users
had been discussed on mailing lists.
Back in April, Brian Behlendorf
had initiated a thread labeled
``Session tracking'' on the \texttt{www-talk} mailing list~\cite{www-talk},
where new ideas for the Web usually got discussed.
Behlendorf wanted a mechanism that would let a server operator
track how a user navigated through a server's pages, but that would
leave control of the identifier with the user.
He proposed that clients send a client-assigned and client-controlled
Session-ID with each request.
This identifier would be constant for the lifetime of a browser
invocation and could be used by a server to identify multiple requests
from the same client.
Thus a server could maintain state if it used this identifier as an
index into a database that tracked a session.\footnote{
Think of a \textit{session} as the interval between when
a server first starts keeping track of a browser and when it stops.
In the context of cookies, a session begins when
a browser first gets a cookie and ends when the browser discards the cookie
(or it expires).}
It could also be used by the server to track a user's traversal
through a web site.

Being somewhat aware of Netscape's cookies,
I proposed a lighter-weight, server-controlled
mechanism that was meant to achieve the same end.
A server could return a single arbitrary datum in a response,
and a client would be obliged to return that datum with each
request to the same server (but not to other servers).
The user could control the degree to which the client participated
in stateful sessions.
As part of the same thread, Lou Montulli
proposed what was essentially Netscape's cookie mechanism.

In August, 1995, I submitted an Internet Draft that
described my Session-ID (later renamed State-Info) proposal~\cite{state-info}.
In response to comments from participants on the HTTP WG
mailing list~\cite{http-wg}, I revised the proposal and submitted a second
I-D in September.
I added sections on privacy to
these early proposals, largely
in response to comments from Koen Holtman who, being Dutch,
gave us a European perspective on privacy issues.\footnote{
Holtman explained to us that European users expect significant
privacy protection.
}

At the December meeting,
with these three related state management
proposals (Behlendorf, Kristol, Montulli)
percolating through the HTTP community,
Larry Masinter, new co-chair of the HTTP Working Group, asked
the interested parties to
form a sub-group to devise a single mechanism and propose it
to the Working Group.
His goal was to separate the issue of state management
from the HTTP specification as a whole, make separate
progress on it, then merge it back into the complete HTTP specification,
thus letting the rest of the working group focus on more urgent issues.

I agreed to lead the state management sub-group, which
began with three people and grew to eight, including
Lou Montulli.
In a triumph of hope over experience, the sub-group
agreed to complete its work by January, 1996, and to publish
its work for the greater Working Group in February, so that
the work could be incorporated into a new HTTP specification
by March, 1996.

In a lively dialog that ensued
on \cite{http-wg} immediately following
the IETF meeting (``making progress on State-Info''),
there seemed to be some
consensus to adopt the State-Info proposal as the basis for HTTP
state management, though there was disagreement about
whether it was better than Netscape's.
(There was also some squabbling about whether IETF process
was being duly followed to determine whether consensus had been achieved.)

The sub-group began its work within days after the IETF
meeting, using teleconferences and a private email list to work out issues.
Although standardizing a new, simpler mechanism was initially more appealing,
we quickly realized that any new mechanism
would have to provide compatibility with Netscape's
already-deployed cookie mechanism.
Adoption of a new, different mechanism would depend
on how quickly such a mechanism could be deployed, both in servers
and in clients, whereas Netscape's mechanism already enjoyed
significant market penetration.
We therefore quickly recognized that it made sense to start from
Netscape's specification.
Having thus shifted our focus, our task became one of tightening
up Montulli's somewhat fuzzy specification.

\subsection{Technical Details: Netscape's Cookies}

So the issues that came under discussion will be clearer,
the following sections describe in more technical detail how
cookies work, according to Netscape's
specification~\cite{NS}.

\subsubsection{Server to Client:  \texttt{Set-Cookie}}

Recall that the earlier, simplified description of cookies
explained that the server sent a \texttt{Set-Cookie} response header
with a cookie value, and the client was expected to return
that value in a \texttt{Cookie} header when it next made a request
to the same server.
In fact, the \texttt{Set-Cookie} header may also include \emph{attribute-value
pairs} that further control cookie behavior.
The (incomplete, informal) syntax for Netscape's \texttt{Set-Cookie} looks like this:

\begin{verbatim}
set-cookie      =       "Set-Cookie:" set-cookie
set-cookie      =       NAME "=" VALUE *(";" set-cookie-av)
NAME            =       attr
VALUE           =       value
set-cookie-av   =       "domain" "=" Domain_Name
                |       "path" "=" Path
                |       "expires" "=" Date
                |       "secure"
\end{verbatim}

\begin{description}
\item{\texttt{NAME=VALUE}} \\
\verb,NAME, is the cookie's name, and \verb,VALUE, is its value.
Thus the header
\verb,Set-Cookie:, \verb,id=waldo,
 sets a cookie with name \texttt{id} and value \texttt{waldo}.
Both the cookie NAME and its VALUE may be any sequence of characters
except semi-colon, comma, or whitespace.

\item{\verb,domain=Domain_Name,} \\
The value for the \texttt{domain} attribute selects the set of servers
to which the client may send the cookie.
By default (if \texttt{domain} is omitted), the cookie may be returned
only to the server from which it came.
If \texttt{domain} is specified, the client may send the cookie to
any host that tail-matches \texttt{domain}, subject to the following
restriction~\cite{NS}:
\begin{quotation}
    Only hosts within the specified domain can set a cookie for a
    domain and domains must have at least two (2) or three (3) periods
    in them to prevent domains of the form: ``.com,'' ``.edu,'' and
    ``va.us".  Any domain that [falls] within one of the seven special top
    level domains listed below only require two periods. Any other
    domain requires at least three. The seven special top level domains
    are: ``COM,'' ``EDU,'' ``NET,'' ``ORG,'' ``GOV,'' ``MIL,'' and ``INT.''
\end{quotation}

The algorithm for matching domains will be discussed frequently and
in more detail later.

\item{\texttt{path=Path}} \\
The value for the \texttt{path} attribute is a URL that specifies
the subset of URLs on the server for which the cookie may
be returned.
If \texttt{path} is specified, the cookie may be returned to any
URL for which \texttt{path} is a string prefix of the request URL.
If \texttt{path} is not specified, ``it as assumed to be the same path as the
document being described by the header which contains the cookie.''

\item{\texttt{expires=Date}} \\
The value for the \texttt{expires} attribute is a timestamp in one of
several standard formats, but in Coordinated Universal Time (UTC, GMT).
If \texttt{expires} is specified, it gives a time after which the
cookie should be discarded.
If not specified, the cookie is supposed to be discarded
``when the user's session ends.''

\item{\texttt{secure}} \\
If this attribute (which takes no value) is present, it means the cookie should
``only be transmitted if the communications channel with the host is a
secure one. Currently this means that secure cookies will only be sent
to HTTPS (HTTP over SSL) servers.''  Otherwise the cookie ``is
considered safe to be sent in the clear over unsecured channels.''
\end{description}

If a client receives a \texttt{Set-Cookie} header for cookie \texttt{NAME},
and the client already has a cookie with the same name, 
\verb,domain,, and \verb,path,, the new information replaces the old.

\subsubsection{Client to Server:  \texttt{Cookie}}

A client is expected to maintain a collection of the cookies
it has received over time.
Before it sends a request to a server, the client
examines all its cookies and
returns zero or more cookies to the server if:
\begin{itemize}
\item{the server's hostname matches the cookie's \texttt{domain} attribute,
according to the rules above; and}
\item{the URL matches the cookie's \texttt{path} attribute,
according to the rules above; and}
\item{the cookie has not reached its expiration time.}
\end{itemize}

The client sends matching cookies in a \texttt{Cookie} request header:

\begin{verbatim}
cookie-header   =       "Cookie:" cookie *(";" cookie)
cookie          =       NAME=VALUE
\end{verbatim}

That is, the cookie name and value are returned.
If more than one cookie matches, they are separated by semi-colon.
\begin{quotation}
[A]ll cookies with a more specific path mapping should be sent
before cookies with less specific path mappings. For example,
a cookie ``name1=foo'' with a path
mapping of ``/'' should be sent after a cookie ``name1=foo2''
with a path mapping of "/bar" if they are
both to be sent.
\end{quotation}

\subsubsection{Commentary}

The Netscape cookie specification seemed to suffer from a number
of syntactic and semantic
deficiencies that the sub-group wanted to fix.

\paragraph*{Syntactic}
\begin{enumerate}
\item \cite{NS} gives no precise syntactic description of the
\verb,Set-Cookie, and \verb,Cookie, headers.
\item{By convention, duplicate HTTP headers can be ``folded'' into
a single header, with the duplicate values separated by comma.
So, for example, if a request has the headers
\begin{verbatim}
	Cookie: cookie1=value1
	Cookie: cookie2=value2
\end{verbatim}
this should be equivalent to
\begin{verbatim}
	Cookie: cookie1=value1, cookie2=value2
\end{verbatim}

However \cite{NS} specifies semi-colon to separate multiple cookies
in \verb,cookie-header,, not comma.
If multiple \verb,Cookie, headers got folded together,
the server might be confused by the comma that
separated cookie values.
}

\item{If comma were to be allowed to separate cookies
according to the HTTP convention, then a ``quoting''
mechanism would be necessary for attribute values, especially \texttt{expires}, for
which two of the acceptable formats contain embedded commas,
and for cookie values (although they were specified to exclude comma).
}

\item{\cite{NS} is vague about what characters are acceptable in the
values of \texttt{domain} and \texttt{path} attributes.
}
\item{\cite{NS} does not specify whether \verb,Set-Cookie, attributes
are case-sensitive or not.}
\end{enumerate}

\paragraph*{Semantic}
\begin{enumerate}
\item{The entire specification seems to lack sufficient precision
to serve as a standard.
For example,
what exactly is the domain matching algorithm?
What exactly is the default value for the \verb,path, attribute?}
\item{There is insufficient discussion of how cookies interact with caches.}
\end{enumerate}

\subsection{State-Info}

Although the sub-group chose not to endorse the State-Info
proposal, many of the ideas that were present in it found
their way into the IETF specification for cookies.
Therefore this section describes some of the important aspects it contained.

The basic State-Info mechanism resembles cookies as I have so
far described them.
A server can send the response header
\verb,State-Info: ,\textit{opaque-information},
with the expectation (or, at least, hope) that the client will
return the same information in a \texttt{State-Info} request header
on the next request to the server.
(The proposal used the same \verb,State-Info, header for both
requests and responses.)
\verb,State-Info,
may \emph{only} be returned to the server from which it came.

Unlike \cite{NS}, the State-Info proposal specifically outlines facilities
that a browser might provide to a user to inspect
and control the state information that it receives.
It also recommends that state information be discarded,
possibly under user control, when a browser exits.
Finally, the State-Info proposal stated that the
\texttt{State-Info} request/response header be passed through proxies
transparently, and that the header should never be cached,
although the associated content could be.

\subsection{Sub-Group Discussions and the First Internet Drafts}

The very earliest mailing list discussions regarding state management
raised issues that recurred throughout the standardization process:
\begin{itemize}
\item{to which domain names a cookie could be forwarded}
\item{privacy}
\item{how cookies interact with proxies}
\end{itemize}

\subsubsection{The \texttt{Domain} Attribute}

\label{domain-attr}

\sloppypar{
The \texttt{domain} attribute of Netscape's proposal quickly became}
a focus of the sub-group's attention.
\cite{NS} provides a primitive kind of domain validation on cookies.
The value of the \texttt{domain} attribute
must tail-match the domain of the request-URI,
subject to a 2-dot/3-dot rule.
As described earlier,
the rules require that, for domain names from
international domains (where the top-level domain is two
characters long, \textit{e.g.}, \texttt{.uk}), the domain name \emph{must}
comprise at least
three components, with a leading `.':  \texttt{.ucl.ac.uk}.
Where the top-level domain is three characters long (\textit{e.g.}, \texttt{.com}),
the domain name \emph{must} comprise at least two dot-separated components:
\texttt{shop.com}.
Thus, a request to \texttt{www.shop.com} may set the \texttt{domain} attribute
to \texttt{www.shop.com} or \texttt{shop.com}, but not \texttt{.com}.

Clearly this mechanism is imperfect as a way of controlling
whether cookies might be sent to unintended destinations.
It relies on the idea that domain names are administered
hierarchically.
It depends on the number of characters in the top-level domain!
On the other hand, it does not require sophisticated
mechanisms such as, say, requiring a server
to send a cryptographically signed list of valid
servers to which cookies may be returned.

The \texttt{domain} attribute actually gives rise to
two kinds of problems, both related to too-wide
sharing of cookies:
1) Cookies may ``leak'' to servers other than those intended by
the originating server.
2) A server in a domain can cause a denial-of-service attack,
either inadvertently or intentionally (``cookie spoofing''),
by sending cookies that will disrupt an application that runs
on another server in the same domain.

\paragraph*{Cookie Leakage}
Suppose a business, \texttt{biz.com}, runs a shop on \texttt{shop.biz.com}.
Suppose that \texttt{shop.biz.com} sets \texttt{domain=biz.com} when it sends
the cookie \texttt{Customer=custid} to a shopper's browser.
\texttt{shop.biz.com} sets \texttt{domain} that way because it wants to share
the cookie with \texttt{pay.biz.com}, the server that handles payment
for shoppers.
So far so good.
Now suppose that \texttt{info.biz.com} is a server that provides
information about \texttt{biz.com}.
If a shopper first visits \texttt{shop.biz.com} and then visits
\texttt{info.biz.com}, the latter site would get the cookie originally
set by \texttt{shop.biz.com}.
That behavior may be undesirable,
either to the administrators of \texttt{shop.biz.com} or
\texttt{info.biz.com}, or to the user, and particularly
if the cookie contains personal information.

\paragraph*{Cookie Spoofing}
Using the same set of servers from the previous example,
suppose that \texttt{info.biz.com} and \texttt{shop.biz.com} are administered
by different parts of \verb,biz.com,.
The administrator of \texttt{info.biz.com} creates an application
that uses cookies, and just happens to use the same cookie,
\texttt{Customer}, but with a value having a different meaning
(\texttt{Customer=otherid}).
Further assume the application \emph{also} sets \texttt{domain=biz.com}.
Now assume the user visits \texttt{shop.biz.com} and then \texttt{info.biz.com}
or \textit{vice versa}.
The second application visited will be
confused by the cookie that had been set by the first application,
because, although each application uses the same named cookie (\texttt{Customer}),
the value for the cookie means different things.
Yet the value of \texttt{domain} causes the cookie to be forwarded
in requests to both servers.
While this scenario could be dismissed as poor design or administration and
is probably inadvertent, an attacker \texttt{attack.biz.com} could
deliberately set cookies to disrupt an application the same way:
using a value for \texttt{domain} that causes wider sending of a cookie,
and choosing a cookie name that is known to be disruptive.

Even at the earliest stages of discussion, we identified privacy and caching
as important and difficult issues to resolve.
The behavior of the Domain attribute and how it affected
the forwarding of cookies, and, thereby, privacy,
was of particular concern, and
the sub-group found it a particularly difficult issue
to resolve satisfactorily.

\subsubsection{Other Issues}
We also discussed how to deal with cookies that come
from servers that ran on different port numbers on the same host,
and concluded they
could share cookies just like two hosts in the same domain.
On the other hand, we quickly adopted the specification for
dealing with caching from my State-Info proposal.

We discussed the security of the cookie contents (should
they be encrypted?), deciding that, because the cookie value was
opaque, a server could, at its discretion, encrypt the cookie value.
We also accepted the idea that users be able to examine and control
cookies, as described in State-Info~\cite{state-info}.

\subsubsection{Preparing for the March, 1996, IETF Meeting}
A late-December conference call continued a discussion that had
emerged on our private mailing list concerning cookies that were,
for example, associated with inline requests for images.
We were concerned about how such cookies might be misused to
track users.
We had stumbled onto what later became a highly contentious issue,
namely ``third-party cookies,'' or ``unverifiable transactions.''
(Section \ref{3rd-party})
Even at this stage we recognized that it would be difficult
to define just exactly what the term meant.

As part of the teleconference,
Lou Montulli agreed that a Netscape technical writer
would create a first I-D for a new specification.
However, the draft we finally received in late January
was unsatisfactorily written.
Our intended February\ 1 deadline appeared to be unattainable.
By mid-February, and still with no follow-up from Netscape,
I took \cite{NS},
folded in elements of my State-Info proposal and comments
that had accumulated on our private mailing list, and produced a
document for discussion by the sub-group.

One part of this draft that got hotly discussed was, surprisingly, caching,
which we thought we had resolved.
At issue was what headers a response would have to contain
to successfully instruct both HTTP/1.0
and HTTP/1.1 caching proxies \emph{not} to cache a \texttt{Set-Cookie} header
along with a response.
We deferred to a separate caching sub-group, which was working on language
that could later be incorporated into the HTTP/1.1 specification.

The draft specification used the same \texttt{Set-Cookie} and \texttt{Cookie} headers
that Netscape used.
We meant for the two mechanisms, the one being defined by us
and the one Netscape was already using, to co-exist.
We expected that HTTP/1.1-compliant
clients and servers would follow our new specification.
In fact, an underlying assumption of our work was that
HTTP/1.1 clients and servers would use ``new''
cookies exclusively, and HTTP/1.0 clients and servers would use only ``old''
cookies, that is, follow \cite{NS}.

In late February, 1996, the sub-group engaged in a hurried
cycle of review and revision of the draft specification,
hoping to beat IETF's cut-off a few days later.
The I-D left a number of questions open, inviting comments
from the HTTP community as well as the state management sub-group.
One of the significant contributions, incorporated
nearly verbatim, was Koen Holtman's wording for ``unverifiable
transactions.''
Holtman had also provided wording to specify the other servers
in a domain with which a client could share cookies.

\subsubsection{Third-Party Cookies}
\label{3rd-party}
``Unverifiable transactions,'' informally
``third-party cookies,'' became a target of much
vituperation and opposition, so it's important to understand
what they are and why their treatment in the I-D became
controversial.

Imagine that you tell your browser to visit \texttt{www.news.com}.
You wouldn't find it particularly surprising when \texttt{www.news.com}
sends your browser a cookie in its response.
If the response from \texttt{www.news.com} included HTML links to images,
including images for advertisements, your browser would ordinarily
load those images automatically.
Suppose those ads came from a third-party site,
such as \texttt{www.ads.com}.
Further suppose that in the responses from \texttt{www.ads.com}, your browser
received cookies from \texttt{www.ads.com}.
If your browser were configured to alert you when you received cookies
(a feature that was not available at that time), you might have been
perplexed about why, having visited \texttt{www.news.com}, you were receiving
cookies from \texttt{www.ads.com}.

Here is how the February, 1996, I-D described third-party cookies:

\begin{quotation}
4.3.5  Sending Cookies in Unverifiable Transactions  A transaction is
verifiable if the user has the option to review the request-URI prior to
its use in the transaction.  A transaction is unverifiable if the user
does not have that option.  User agents typically use unverifiable
transactions when they automatically get inlined or embedded entities or
when they resolve redirection (3xx) responses.  Typically the origin
transaction, the transaction that the user initiates, is verifiable, and
that transaction directly or indirectly causes the user agent to make
unverifiable transactions.
\end{quotation}

Third-party cookies raised a number of concerns,
first among them the ``surprise factor.''
You receive cookies from sites you were unaware you visited.
This concern was evidenced in news and Usenet articles where
people examined the ``cookie file'' on their machine and found
entries from sites they didn't recognize.
This led them to fear that their machine's security
had somehow been violated,
although no ``break-in'' had actually occurred.
Even when they understood what happened,
some users felt that
a web site had stored information on their machine
without permission.

A second concern was that users had virtually no control over the process.
Short of turning off automatic image loading, a user could not avoid
receiving third party cookies.
The transaction between the user's browser and the third-party site
(\textit{e.g.}, \texttt{www.ads.com}) was \emph{unverified}, in the sense that the user
had very little means to tell \emph{in advance} that their visit to \texttt{www.news.com}
would lead to visits to \texttt{www.ads.com}.
The I-D therefore called for browsers to provide a means, however
primitive, for users to be able to predict when third-party transactions
might take place, as well as a way ``to determine whether
a stateful session is in progress.''
It further stipulated that browsers \emph{must not} accept cookies
from, or send cookies to, third party sites, although the browser could offer
an option that permitted them, provided its default setting was to
disallow them.

\begin{quotation}
When it makes an unverifiable transaction, a user agent must only enable
the cookie functionality, that is the sending of Cookie request headers
and the processing of any Set-Cookie response headers, if a cookie with
a domain attribute D was sent or received in its
\label{origin-transaction}
origin transaction,
such that the host name in the Request-URI of the unverifiable
transaction domain-matches D.

User agents may offer configurable options that allow the user agent, or
any autonomous programs that the user agent executes, to ignore the
above rule,
\textbf{
so long as these override options default to ``off.''}
[emphasis added]

NB:  Many current user agents already provide an acceptable review
option that would render many links verifiable.
\begin{itemize}
\item When the mouse is over a link, the user agents display the link
 that would be followed.

\item By letting a user view source, or save and examine source, the user
 agents let a user examine the link that would be followed when a
 form's submit button is selected.

\item When automatic image loading is disabled, users can review the
 links that correspond to images by using one of the above
 techniques.  However, if automatic image loading is enabled, those
 links would be unverifiable.
 \end{itemize}
\end{quotation}

Apart from surprise and control, what's the big deal about third-party
cookies, anyway?
The I-D said (with respect to the unverifiable transaction rule above):

\begin{quotation}
The [above] rule eases the implementation of user agent mechanisms
that give the user control over sessions\ldots\  by
restricting the number of cases where control is needed.  The rule
prevents malicious service authors from using unverifiable transactions
to induce a user agent to start or continue a session with a server in a
different domain.  The starting or continuation of such sessions could
be contrary to the privacy expectations of the user, and could also be a
security problem.
\end{quotation}

In other words, although most users may not initially
be aware of it,
third-party cookies raise serious privacy concerns
because they facilitate the ability to build profiles.
Suppose that \texttt{www.ads.com} serves ads not just for \texttt{www.news.com}
but for a large number of other content web sites as well.
Each time one of these content sites returns a page to your browser,
it may contain links to \texttt{www.ads.com}.
Each time it requests an ad from \texttt{www.ads.com},
your browser returns the \texttt{ads.com} cookie.
The HTTP \texttt{Referer} [sic] header that the browser normally returns
with the request will contain the URL of the page in which the \texttt{www.ads.com}
link appeared.\footnote{
The advertising sites can also have an agreement with the content sites
whereby the latter embed ``useful'' information in the URLs
for the former, obviating the need for
\texttt{Referer}.}
Thus \texttt{ads.com} can accumulate a profile of the web sites you (anonymously)
have visited.
That is, without specifically knowing who ``you'' are, it will know
what set of web sites you have visited.
Thus \texttt{ads.com} can serve ads that it infers might interest
you, or it can show you a particular ad for a set number of times.
Even more worrisome is the potential for \texttt{ads.com} to
share or sell profiles to other companies.

The notion that they are being watched
as they browse the Web troubles some people.
It's a bit like walking into a store and having someone follow you
around, keeping track of what you look at.
Moreover, if you return to the store on another occasion, the
tracking resumes.
The observer who follows you around might even make suggestions
about things you might want to buy, based on the set of things
you've looked at so far.
Now suppose you make a purchase using a credit card.
All of the observations that have been made so far can potentially
be linked to ``you'':
your name and address, your likely income, based on your ZIP code,
what kind of car you own, what bank accounts you have, and so on.

Whereas there is concern that a single entity, say \texttt{www.shop.com},
may accumulate a profile about you, particularly once you've made a
purchase there,
you presumably have some kind of trust relationship with them.
On the other hand, you have no such relationship with the
advertising sites that can accumulate a much wider profile.
Moreover, one can argue that their commitment is more likely to be to the
companies who use their service, rather than to the
consumers they track, and the advertising networks are therefore more likely
to share profile information with their customers.

Strangely enough, when we added the words about ``unverifiable
transactions'' to the I-D, our direct motivation was not
advertising networks (which at best we were only dimly aware of at
that time).
Instead, Koen Holtman had independently discovered the
theoretical potential to use third-party cookies for profiling, and he
persuaded members of the sub-group that Europeans, at least,
would be very troubled by the potential abuse of privacy
they could promote.
In fact,
in October, 1995, the European Union (EU) issued a directive
on the processing of personal data~\cite{95/46/EC}
that called for informed consent of users before such
data could be collected, which would appear to
render unverifiable transactions illegal under EU law.
In any event, many of us expected that if users' privacy
expectations were violated by third-party cookies,
their comfort with this new medium would be diminished,
probably to its detriment.

\subsection{Resolving Outstanding Technical Issues}
At the March, 1996, IETF meeting, I presented the State Management
I-D at the HTTP Working Group session.
The outstanding issues were:
\begin{itemize}
\item{In the syntax of the Set-Cookie header, were spaces allowed around the `='
signs?}
\item{What is the default for the Path attribute when none is specified?
(Do we truly treat the URL as a \emph{prefix}, or do we remove the last
path element?)}
\item{In the \texttt{Cookie} header, what rules apply
for domain-matching to decide which cookies to return with a request?}
\item{What precise ordering rules apply when multiple cookies
can be returned with a request?}
\item{How can we get ``old'' and ``new'' cookies to interoperate?}
\item{How can we deal with older caching proxies?}
\end{itemize}

In subsequent discussions on the sub-group mailing list, we identified
three additional issues:
\begin{itemize}
\item{How should we address cookie ``spoofing'' (described earlier)?}
\item{How should we deal with the fact that existing implementations
let the \texttt{Domain} attribute's value start with `.'? }
\item{Could we change the separator between cookies from ; to ,?}
\end{itemize}

Over the next three weeks, in response to conference calls and
sub-group discussion, I prepared a series of private drafts to
address these questions.
The most important results were:
\begin{description}
\item [Interoperation]
We added a \verb,Version, attribute to the \verb,Set-Cookie, header to describe
what version of cookie it is, to distinguish ``old'' from ``new.''.
\item [Cookie Spoofing]
We added attributes to the \verb,Cookie, header so an application could
tell whether a given cookie belonged to it.
Initially these attributes were \verb,Cookie-Domain,, \verb,Cookie-Path,, and
\verb,Cookie-Version,,
but we shortened them to \verb,$Domain,, \verb,$Path,, and \verb,$Version,.
(We were still unsure, however, whether existing applications
would tolerate the \verb,$, in the \verb,Cookie, header.
Lou Montulli guessed \verb,$, would have no adverse effect.)
\item [Domain Matching]
We changed the wording to have the effect of allowing
cookie sharing with at most one level of additional domain name
hierarchy.\footnote{
If \texttt{Domain=.example.com}, the browser may accept cookies from,
and send cookies to, \texttt{b.example.com} but not \texttt{a.b.example.com}.
}
Note that, while this is more restrictive than \cite{NS},
we felt that providing a single level of hierarchy struck a proper balance
between functionality and security.
\end{description}

The resulting draft emerged for
full review as \texttt{draft-}\texttt{ietf-}\texttt{http-}\texttt{state-} \texttt{mgmt-01}\footnote{
From the web page
\url{http://portal.research.bell-labs.com/~dmk/cookie.html}
you can find the complete set of publicly available drafts
of the cookie specification, often with change bars
between newer and older versions.}
on April 26, 1996.
Thereafter, until June 10, there were almost no comments.
I submitted a quick revision, \texttt{state-mgmt-02}~\footnote{
In the names of further I-Ds, the \texttt{draft-ietf-http} prefix is omitted.
}, because
Roy Fielding noticed that the previous draft contained change marks
that would be unacceptable in an RFC.

\subsection{Working Group Last Call: June 10, 1996}

Larry Masinter made a working group
Last Call to move the I-D to Proposed Standard,
which sparked more discussion on \cite{http-wg}, particularly
regarding cookie sharing.
Marc Solomon asked whether the specification could be
changed so a server could enumerate a set of servers to
which it wanted a cookie returned.
The answer was that such a mechanism makes cookie sharing
too easy.
That response provoked an exchange between Koen Holtman
and Benjamin Franz.
Holtman said it was important to make it relatively hard
to share cookies in ways other than those that the specification
allows.
Franz retorted that there are many covert ways to share
cookie information, and that if the specification did not
provide legitimate ways to do so, ways that would be evident
to inspection, web sites would employ other methods that were
harder to detect and would thus provide a false sense of security.
However, this discussion ultimately resulted in no change in the I-D.

More editorial changes led to \texttt{state-mgmt-03},
announced on July 16.
Changes from the previous I-D included:
\begin{itemize}
\item{a new \texttt{Comment} attribute.
One concern had long been that there was no way
for a user to know what a cookie was being used for,
because the value of a cookie is ``opaque.''
The purpose of \texttt{Comment} is to let servers describe
(briefly) what a cookie is for.}
\item{more explicit information about which HTTP headers
had to be used with cookies to produce whatever caching
behavior was desired, courtesy of the HTTP caching sub-group.}
\item{clarification that \texttt{Set-Cookie} can accompany any HTTP response,
and \texttt{Cookie} can accompany any HTTP request.}
\end{itemize}

\subsection{Becoming an RFC}

On July 23, Larry Masinter formally asked the IESG to publish
\texttt{state-mgmt-03} as a Proposed Standard.
On August 6, the IESG made its Last Call for the I-D.
A week later, Paul Grand
addressed comments
to \cite{http-wg} regarding the cookie specification, identifying
two concerns:
\begin{enumerate}
\item{the requirement that by default user agents should delete
all state information when they terminate.}
\item{the requirement that by default unverifiable transactions
be disabled:
\begin{quotation}
In section 4.3.5 eliminating the ability of having ``unverifiable''
redirection impairs the ability of the web service (chosen by the user agent
operator) to engage in using the services of a third party for advertising,
content building, download specialized ``plugins'' or other usage.  This hurts
web commerce.  Why is this proposed?
\end{quotation}}
\end{enumerate}

I responded that both these requirements adopted the attitude that
the user should get to decide what happens to the state information,
and that ``informed consent'' should be a guiding principle.
Harald Alvestrand, responding on the IESG's mailing list\footnote{
Alvestrand was, at the time, Applications Area Director, and thus a
member of the IESG.
Grand had addressed his comments to the IESG's mailing list
(which is not publicly archived),
and Alvestrand responded to him there.
Whether the response was an ``official'' IESG response or just
an Area Director's response is unclear.},
said:
\begin{quotation}
It seems to me that the draft is heavily emphasizing the right of the
user to be aware of which servers may be tracking his state; I cannot
see that the ``third party services'' you refer to are significantly hurt
by being unable to track the user's state in relation to that third party.
\end{quotation}

In other words, there was support in the IESG for the wording about a
browser's default behavior regarding state information and a dismissal
of the concern that such a default would impair advertising networks.

\subsubsection{IESG Comments, Approval}

After that brief exchange, which seemed to settle the issue,
at least so far as the IESG was concerned, there was no further discussion
regarding state management for two months.
Finally, on October 15 the IESG made comments
that needed to be addressed before the specification could
become a Proposed Standard.
The IESG was concerned that the ``view source'' feature of
browsers might be an inadequate
means to check for unverifiable transactions.
I explained that that section of the specification was a compromise
to avoid placing user interface burdens on browser implementors,
and I left it to the Applications Area Director, Keith Moore, to
defend that position.

A private discussion among Moore, Masinter, and me
considered whether having a user examine HTML \textit{via} ``view source''
was a minimally acceptable means of previewing potential
unverifiable transactions.
After some back and forth, we crafted some words that expressed
the intent of the state management sub-group while simultaneously
satisfying the IESG's concerns:
\begin{quotation}
Many user agents provide the option for a user to view the HTML source
of a document, or to save the source to an external file where it can be
viewed by another application.  While such an option does provide a
crude review mechanism, some users might not consider it acceptable for
this purpose.  An example of a more acceptable review mechanism for a
form submit button would be that, when the mouse is over that button,
the user agent displays the action that would be taken if the user were
to select the button.
\end{quotation}

This formulation attempts to strike a balance between imposing
a detailed user interface requirement on browser vendors and giving them
no guidance at all about what kind of information the browser should provide
a user.
It says that, ideally, a browser should provide a user with obvious
feedback about where a link will take the user.
However, at a minimum, a sophisticated user would be able to look
at the (HTML) source of a page to decide where the link goes.

Internet Draft \texttt{state-mgmt-04}, which incorporated
these changes to satisfy the IESG's concerns was announced on November 4.

On November 21, the IESG requested
another small change regarding the definitions of
``fully-qualified host name'' and ``fully-qualified domain name.''
I submitted the new changes as \texttt{state-mgmt-05}, which
was announced on November 22.
The IESG approved the I-D on December 2,
which meant it would be published as a Proposed Standard RFC.

\subsection{A Compatibility Issue Surfaces}

Meanwhile, on October 29, Koen Holtman had alerted me to another
issue that would return in greater force later.
In an experiment to test interoperation of ``old'' and
``new'' cookies, he found that Microsoft Internet Explorer (MSIE) Version 3
and Netscape Navigator Version 3 behaved differently
when they received ``new cookies.''
He had a server send the following ``new'' cookie to an MSIE v3 client:

\begin{verbatim}
Set-cookie: xx="1=2\&3-4";
    Comment="blah";
    Version=1; Max-Age=15552000; Path=/;
    Expires=Sun, 27 Apr 1997 01:16:23 GMT
\end{verbatim}

When MSIE sent the cookie back to the server, it looked
like this:
\verb,Cookie:, \verb,Max-Age=15552000,, whereas Navigator
sent \verb,Cookie: xx="1=2\&3-4",.
I also verified the behavior with MSIE version 2.
Clearly two common ``old cookie'' browsers behaved differently
when they received ``new'' cookies, a situation that would
be unacceptable long-term.

As luck would have it, I was at the December, 1996, IETF meeting
when Holtman coincidentally reminded me of the problem by email.
I described it (as an ``MSIE bug'') to Yaron Goland of Microsoft,
and he promised to look into it, and that
any problem found would be addressed in version 4 of MSIE.

\subsubsection{Errata for the forthcoming RFC}

On December 31, Holtman sent a message to \cite{http-wg} to report
the above problem, and a second one having to do with incorrect
information in the I-D concerning headers that affect caching.
At this point I began to assemble a list of errata.
A procedural question arose about whether to wait for
the approved I-D to emerge as an RFC first, or to produce
a new Internet Draft that corrected the observed problems
and try to replace the approved I-D in the RFC Editor's queue.
After discussion, Larry Masinter and I decided to wait
until the RFC appeared, which I assumed would not take very long.

At Masinter's urging, I wrote up the errata to
the putative RFC, and on February 4, 1997, the errata I-D
(\verb,state-mgmt-errata-00.txt,) was announced to \cite{http-wg}.
It included the aforementioned correction for cache headers,
remarks about the observed MSIE problem, and minor typographical
corrections.
Discussion on the HTTP-WG mailing list led to further minor
wording changes.

On February 14, Masinter changed his mind in light
of the seriousness of the compatibility issue and
urged me to put together a revised
I-D of the entire cookie specification
to supersede the one pending in the RFC Editor's queue.
He didn't want an I-D with known flaws to become an RFC if we
could help it.
On February 17, I created such a draft for inspection by the
HTTP community before I submitted it to the Internet Drafts Administrator
two days later.
Foteos Macrides responded by describing
how several existing cookie implementations
failed when servers sent them ``new cookies,''
confirming that further revisions were in order.
Ironically, events overtook our plan:
RFC 2109~\cite{rfc2109},
\textit{HTTP State Management Mechanism}, was announced on
February 18.

\section{After RFC 2109: February, 1997 to October, 2000}
\label{history-2}
Recall that ``RFC'' stands for ``request for comments.''
\emph{This} RFC certainly produced comments.
First to weigh in was Yaron Goland, who,
while commenting on the new draft I had planned to submit,
by extension was commenting on RFC 2109 as well.
He contended that the entire mechanism to support ``old'' and ``new''
cookies
with the same set of headers depended on how Navigator
``handles illegally formatted cookies.''
(Hereafter I'll refer to v0 cookies to refer to those
compatible with \cite{NS}, and I'll refer to v1 cookies as
those that conform to RFC 2109.)
He also (justifiably) criticized the draft's words that,
in identifying an incompatibility with MSIE, accused MSIE of
``send[ing] back the wrong cookie name and value.''

\subsection{Fixing the Incompatibility}
\label{incompat}
\subsubsection{The Problem}

Here's the technical issue.
We want the four possible combinations of client and server to interoperate:
\begin{itemize}
\item v0 client, v0 server (\texttt{Cv0Sv0})
\item v0 client, v1 server (\texttt{Cv0Sv1})
\item v1 client, v0 server (\texttt{Cv1Sv0})
\item v1 client, v1 server (\texttt{Cv1Sv1})
\end{itemize}
Recall the earlier example that led to our first recognizing
a problem:
\begin{verbatim}
Set-cookie: xx="1=2\&3-4";
    Comment="blah";
    Version=1; Max-Age=15552000; Path=/;
    Expires=Sun, 27 Apr 1997 01:16:23 GMT
\end{verbatim}
The observed failure was for the \texttt{Cv0Sv1} combination and depended on
how a v0 client interpreted unrecognized attributes.
A v0 client would understand \texttt{Path} and \texttt{Expires}, but
not \texttt{Comment}, \texttt{Version}, or \texttt{Max-Age}, which
are new to v1.
We incorrectly assumed that all clients would treat the
first attribute name-value pair in a \texttt{Set-Cookie}
header as the cookie's name and value; in fact, that's what Navigator did.
But \cite{NS} did not specify what the client should do under
these circumstances; MSIE treated the \emph{last}
unrecognized attribute-value pair as the cookie's name and value.
Given that MSIE had a 20-30\% (and growing) share of the browser
market at the time, it was clear that RFC 2109 had to be revised;
the compatibility issue was too serious.
Thus the discussions began that aimed at resolving the
v0/v1 compatibility problem.

\subsubsection{Independent Header Proposal}

The most obvious solution was to have two parallel sets of headers.
That is, a server would send both v0 and v1 cookies in responses.
The solution was so obvious, in fact, that the state management sub-group
had already considered and rejected it a year earlier.
The problem with parallel independent headers is that there's a ``chicken-and-egg''
deployment problem.
Until the population of clients that understand
v1 cookies is sufficiently large,
there's no incentive for a web site to use them.
And since browser vendors would likely support v0 cookies
for some time, web sites could continue to send only v0 cookies
without fear of loss of functionality.
Even a well-meaning site would have to send two sets of
cookie headers until it concluded ``enough'' clients understood v1 cookies,
after which it could cease to send v0 cookies.
Any site that sent both headers would incur extra bandwidth expense.

Following some discussion on the sub-group mailing list,
I prepared an I-D that essentially created
a new mechanism with two new headers, \texttt{Set-Cookie2} and \texttt{Cookie2},
nearly independent of v0 cookies,
and I solicited comments.
A client that understood only v0 cookies would ignore the
\texttt{Set-Cookie2} header and would return v0 cookies as before.
Only when a server sent v1 cookies to a v1-capable client
would a client send v1 cookies.\footnote{
HTTP clients and servers are supposed to ignore headers they
do not recognize.
}

Because the incompatibility we found opened an opportunity
to add other desirable features to the specification,
the revised I-D also included a new attribute, \texttt{Discard},
that had been suggested earlier, and that had been well-received
in discussions.
``The Discard attribute instructs the user agent to
discard the cookie unconditionally when the user agent terminates.''
\verb,Discard, overrides \verb,Max-Age,.

However, my describing a mechanism that was independent of
current practice provoked Larry Masinter to question
why exactly we would continue to call them ``cookies,''
and why we should be in a hurry to revise RFC 2109 this
way if there was no installed base?
In effect we would be adding an entirely new feature to HTTP.
(The IETF emphasizes ``rough consensus and running code,''
and there was little running code.)
But, the discussion that ensued pointed out that we were trying
to fix an interoperation bug in RFC 2109.

\subsubsection{Duplicated Cookie Value Proposal}

As a way to make minimal changes to RFC 2109,
Jeff Mogul proposed a clever way to avoid two headers.
Using the previous example, he proposed that servers send
the cookie name and value twice, once at the beginning
and once at the end:
\begin{verbatim}
Set-cookie: xx="1=2\&3-4";
     Comment="blah";
     Version=1; Max-Age=15552000; Path=/;
     Expires=Sun, 27 Apr 1997 01:16:23 GMT;
     xx="1=2\&3-4"
\end{verbatim}

Clients like Navigator, which interpret the first unrecognized
attribute-value pair as the cookie name and value, and
those like MSIE, which treat the \textit{last} such attribute-value pair
as the cookie name and value, would both get the ``right answer.''
The rule that ``If an attribute appears more than once in a
cookie, the behavior is undefined'' would be amended to say
``unless the value
is the same in each appearance, in which case subsequent
appearances are [ignored].''

\subsubsection{Additive Proposal}

A discussion about the proper resolution of this dilemma
developed on the state management sub-group mailing list.
Goland objected to the two-value solution
because of the extra overhead of the duplicated cookie header:
the server would have to send two copies of all the cookie information,
v0- and v1-style.
Instead, he proposed a variant of the duplicate header solution in my draft.
In contrast to my ``parallel proposal,''
Goland proposed what we began to call the ``additive proposal.''
A server would send \texttt{Set-Cookie} as before.
However, if it understood v1 cookies, it would also send
a \texttt{Set-Cookie-V1} header that contained the attribute-value
pairs that were new for v1 cookies.
Then the (v1-capable) client would form a complete cookie
by combining corresponding pieces
from the \texttt{Set-Cookie} and \texttt{Set-Cookie-V1} headers.
Goland did not propose a new, matching \texttt{Cookie-V2} header.

Marc Hedlund noted that the ``extra overhead'' of Mogul's two-value
solution depended on the length of the cookie name and value and
might not be much greater than Goland's additive header proposal, unless
the cookie name or value is long.
Goland asserted that, in his experience, cookies tend to be long.
Moreover, I discovered, upon doing a simple experiment, that Mogul's
solution would not work for MSIE v2, so that possible solution
was dropped.

In a separate message to the sub-group, Goland provided comments
to my (privately available) draft and expressed
unhappiness with the description of cookie lifetime (\texttt{Discard},
\texttt{Max-Age}), domain-matching, and requirements on user agents to allow
a user to control and inspect cookies.
I explained (because Goland had not been following those discussions
at the time) that the words in RFC 2109 had been arrived at through
hard-won consensus and had been considered of high importance by
the sub-group.

On March 5, I made available to the sub-group a draft that incorporated
the additive two-header solution to the RFC 2109 compatibility bug.
Hedlund said he was concerned about the complexity of grouping
components from two separate headers to build a single cookie.
Goland felt that the new draft satisfied his compatibility concerns,
though he restated his unhappiness about the other issues mentioned
above, and he added that the \texttt{Secure} attribute's description was
``fuzzy.''

\subsection{Other Issues}

In addition to the compatibility problem we were already
wrestling with, a number of issues re-surfaced that we thought
had been resolved,
\begin{enumerate}
\item unverifiable transactions, out of which grew the
idea of ``certified cookies''
\item domain-matching rules, and \verb,Port,
\item the \verb,Comment, attribute
\end{enumerate}

\subsubsection{Unverifiable Transactions, Revisited}
On March 13, Dwight Merriman of DoubleClick expressed his
opposition to the specified default behavior, saying,
``[d]isabling stateful sessions for unverifiable transactions by default is
basically equivalent to not allowing them at all, because 99\% of the
population will see no reason to change the default,''
and he went on to describe how the default would have a negative
effect on advertising networks.
He did concede ``that privacy is a concern and an important issue.''
Goland asserted that the default
could hurt smaller web sites that rely on advertising
for support, and that their demise would reduce the Web's diversity.
Others pointed out that the default setting would by no means disable
advertising networks, but it would affect
business models that depended on third-party cookies.

The ensuing wide-ranging
discussion on \cite{http-wg} observed that ``cookie sharing'' is possible
by other means, and that the ``unverifiable transactions''
rule does not prevent it.
Dan Jaye
stated
that, while there might be the potential for privacy abuse,
``the number of web sites and applications that make use of
`unverifiable transactions' for legitimate, non-privacy invading uses is
significant and growing.''~\footnote{
Jaye went on to propose a ``certified'' or ``trusted'' cookies,
about which more later.
}
However, Marc Hedlund said ``the concern of the state management subgroup was
crafting a specification that did not create \emph{new} privacy problems.''
And Koen Holtman added,
``The key words in the cookie spec are `privacy expectations of the
user'.  The spec does not really claim to raise the level of privacy
on the web, it claims to remove some behaviour which is `contrary to
the privacy expectations of the user'.''

After 69 messages on the topic in
seven days, Larry Masinter
called for an end to the discussion
(because it was distracting from work on HTTP/1.1 proper)
and invited those who supported a change
in the ``unverifiable transaction'' default to 
write an I-D that outlined their version of the protocol.

A few days later, Dan Jaye posted a proposed revised section
on unverifiable transactions that called for the user agent
to ``verify that the request-URI comes from a trusted
domain by placing a request to a certificate authority to get the
credentials of the domain.''
Although this posting was inadequate because of its
lack of specificity, it did contain the kernel of an idea
that Jaye was encouraged to elaborate.
He did so a few days later, although he focussed more
on certifying the \emph{identity} of the sender of a
cookie than the sender's intended \emph{use} of the information collected.

On March 18, I submitted as a new I-D a revision
of the draft we had been discussing,
which reflected the additive two-header proposal to solve
the incompatibilities we had found, and which
added the \verb,Discard, and \verb,Comment, attributes.
(The I-D got named \verb,state-man-mec-00,
because IETF naming rules precluded calling it \verb,state-mgmt-06,.)

\subsubsection{Domain-matching, again, and \texttt{Port}}
In addition to the incompatibility described above,
Goland again raised the domain-matching rules as an issue,
and he questioned why a cookie may be
returned only to a server with the same port number as
the server from which the cookie arrived.
He also objected to the parts of the specification
that placed requirements on the user interface, claiming
it was out of bounds for the IETF to do so.

The ``which port'' issue opened a discussion about desirable
behavior with respect to port numbers and cookies.
I initially proposed dropping any port restrictions;
a browser could return a cookie to any port on any server
that otherwise met the domain-matching rules.
However, that seemed to open a potential security hole,
where cookies sent by a server on one port could leak
to other servers on the same host but running on different
ports.

Over the next few days we stumbled into a consensus
to add the \verb,Port, attribute to solve the perceived
problem.
\verb,Port, would behave as follows.
If a server sends a cookie without a \verb,Port, attribute,
a client may return the cookie to a server running
on \emph{any} port on the same host.
If \verb,Port, is specified without a value, the client
may return the cookie only to the same port from which
it was received.
If \verb,Port, has a value, the value must be a comma-separated
list of valid ports to which the cookie may be returned,
and the sending port must be one of them.
Thus \verb!Port="80,443"! would direct a client to send
the cookie to servers on either port 80 or port 443.

\subsubsection{\texttt{Comment}, and \texttt{CommentURL}}
\label{commenturl}
Among his extensive set of comments,
Goland said that \verb,Comment, needed to
be tagged as to which language it used.
(Internationalization, or \emph{i18n}, was and remains a
hot topic in the IETF.)

Jonathan Stark proposed
yet another new attribute, \verb,CommentURL,
that would resemble \verb,Comment,,
except its value would be a URL that a user
could inspect to understand the cookie's purpose.
\verb,CommentURL, was an attractive idea because it
could direct a user to much more information than
\verb,Comment, could convey.
Moreover, the URL could finesse the language issue of
\verb,Comment, by relying on HTTP's language negotiation
capabilities to provide useful information to a user
in her native language.
Finally, the page associated with \verb,CommentURL,
could explain, at the point where a user must decide
whether or not to accept a cookie, what the cookie is for.
Further discussion identified some potential issues:
\begin{itemize}
\item What happens if the response to accessing the \verb,CommentURL,
page \emph{itself} returns a cookie?
\item How, exactly, should \verb,CommentURL, work?
Ideally a user should be able to examine the
\verb,CommentURL, information \emph{before} accepting a cookie.
Should a browser pop open a window with the information automatically?
(This could cause a loop if the \verb,CommentURL, page also
included a cookie.)
\item Should the content type of the \verb,CommentURL, page
be restricted?
Suppose it points to executable code?
\item Should there be a prescribed relationship between the domain
of the \verb,Request-URI, that gave rise to the cookie and
the domain of \verb,CommentURL,?
In other words, may I get a cookie from \verb,www.a.com, with
a \verb,CommentURL="http://www.b.com",?
\end{itemize}

We did not resolve these issues until after the next IETF meeting.

\subsection{Memphis IETF Meeting: April, 1997}

The flurry of activity that led up to the IETF meeting in Memphis
in early April, 1997, was followed by a lull.
At the meeting I laid out the two areas of contention,
compatibility and default user-agent behavior.
I also mentioned that Dan Jaye was working on a proposal
for ``certified cookies'' that might break the impasse
about unverifiable transactions.

Because the HTTP-WG as a whole was focussed on completing
the larger HTTP specification, the decision was made
to move the distracting and contentious
state management discussions to a new, separate
mailing list, HTTP-STATE~\cite{http-state}.
The goal was to try to reach consensus on the new
list and then bring the result back to HTTP-WG.
(Procedurally, \cite{http-wg} remained the mailing
list of record for reaching working group consensus
on HTTP state management.)

\subsubsection{Certified Cookies}

Jaye's ``certified cookie'' idea had evolved and looked like it might
resolve the ``unverifiable
transaction'' default setting issue.
The idea was to create a mechanism
whereby the user can configure her browser to accept
cookies, even third-party cookies,
from senders that have been pre-authorized.
The cookie senders would obtain a cryptographic
certificate that attests to the sender's identity and
that asserts how they would
use the information collected \textit{via} cookies.
They would send the certificate along with the cookie.
The browser could then verify the certificate's authenticity
and check whether the uses the sender would make of the cookie information
would fall within the bounds the user has configured.
If so, the cookie would be accepted without further
notification to the user.
If not, either the cookie could be rejected outright, or
the user could be asked whether to accept the cookie.
Separate (private) agencies would audit
the behavior of the organizations that obtain
cookie certificates to verify that the information
collected \textit{via} cookies was indeed being used as
the senders claimed.

Certified cookies promised some nice properties.
\begin{itemize}
\item Users could fine-tune what cookies they're willing to accept,
based on how the information collected would be used.
\item Web sites that send cookies could allay users' fears about
how information would be used by the way they labeled cookies
and by their willingness to have that use audited.
\end{itemize}

On May 15, Dan Jaye's first I-D, \verb,jaye-trust-state-00,,
was announced.
The mechanism proposed there for ``certified cookies'' was
an extension to RFC 2109+\footnote{
I'll use the notation ``RFC 2109+'' to refer to the RFC that
we were working on to supersede RFC 2109.}.
I encouraged Jaye
to design it as an add-on to the cookie
specification, rather than to try to merge the two.
In late May, there was some discussion on \cite{http-state}\footnote{
``Advertisers win one in debate over `cookies'\,''
}
about whether this was the correct approach, or
whether merging the two would be better.
The consensus was that RFC 2109+ without Jaye's extensions
was necessary as a base for those situations where no
certified cookie was present.
Moreover, it seemed likely that agreement could be reached
on it sooner than on Jaye's wholly new proposal.

\subsubsection{Trying to Achieve New Consensus:  May-July, 1997}
In early May, 1997, I submitted \verb,state-man-mec-01,, which
included the new \verb,Port, attribute,
but not \verb,CommentURL,, because there was no
consensus for it.
Despite the discussions concerning them,
the language regarding ''unverifiable transactions'' remained unchanged.
This new draft sparked virtually no discussion (because of ``cookie fatigue''?).
However, a stray thought led me to reexamine the
wording for ``unverifiable transactions,'' and to
start a private discussion with Koen Holtman about remedies.

It turned out that the wording in RFC 2109 was even more
restrictive regarding ``unverifiable transactions''
than we intended or that even the fiercest opponents
of the RFC had accused us of.
The wording quoted in Section~\ref{origin-transaction}
regarding the default setting
implies that a session cannot be initiated
\textit{via} an unverifiable transaction (in
addition to all the other restrictions)
unless the ``origin transaction'' resulted in a cookie's being returned
to the client.
In other words, if the client did not receive a cookie with the
origin response, it could \emph{never} accept cookies for
responses for, for example, embedded images.
We massaged the words (and introduced the concepts of \textit{reach} and
\textit{third-party host})
to remove the above error without changing our otherwise intended behavior
with respect to ``unverifiable transactions.''
In mid-June, \verb,state-man-mec-02, was issued
to correct the error.

And then there was silence on \cite{http-state}.
Inasmuch as silence could be construed as
indifference, acceptance, or lack of awareness,
I asked Larry Masinter to issue a Working Group
Last Call, hoping we could pass the I-D to the IESG,
and he did so on July 8.
The Last Call once again brought forth comments,
and the volume eventually led Larry Masinter to
withdraw the Last Call.
The comments focussed on three ever-popular issues: the domain
matching rules, \verb,CommentURL,, and the rules
for combining \verb,Set-Cookie, and \verb,Set-Cookie2,.

There was also some discussion about whether it
made sense to continue discussion of RFC 2109+ at all.
There were suggestions to take RFC 2109 off the
IETF standards track and mark it either ``Experimental''
or ``Historical.''
However, doing nothing was unacceptable, because of the
acknowledged technical flaws in RFC 2109.
There was also a request to ``document how cookies are implemented today.''
Indeed such a document would be useful,
but it would be
completely separate from RFC 2109+.
I was not willing to write it, however, and I thought it could
be written most effectively by the browser vendors, but
no one volunteered to do so.

\subsubsection{Domain-Matching}
Dave Morris noted some deficiencies in the domain-matching
wording that would adversely affect \emph{intra}nets.
We crafted some words that would allow cookies to work
as intended even if a domain name was not fully qualified.

\subsubsection{\texttt{CommentURL}}
Dave Morris objected to the fact that, despite a high degree
of support for its addition, \verb,CommentURL, was absent
from \verb,state-man-mec-02,.
At one point during the ensuing discussion,
he said, ``It would be irresponsible protocol design to not provide
the more complete approach [than \verb,Comment,] in the protocol.''
I agreed that there was support for it, but that we had not
worked out words that described how a user agent should deal
with receiving or sending cookies while inspecting the
\verb,CommentURL,.
The resulting discussion (``Removing CommentURL''~\cite{http-wg})
produced further vigorous support for the addition of
\verb,CommentURL,, as well as these words to address cookies
within \verb,CommentURL,:

\begin{quotation}
The cookie inspection user interface may include a facility whereby a
user can decide, at the time the user agent receives the \verb,Set-Cookie2,
response header, whether or not to accept the cookie.  A potentially
confusing situation could arise if the following sequence occurs:

\begin{itemize}
\item the user agent receives a cookie that contains a \verb,CommentURL,
attribute;

\item the user agent's cookie inspection interface is configured so that
it presents a dialog to the user before the user agent accepts the
cookie;

\item the dialog allows the user to follow the \verb,CommentURL, link when the
user agent receives the cookie; and,

\item when the user follows the \verb,CommentURL, link, the origin server (or
another server, \textit{via} other links in the returned content) returns
another cookie.
\end{itemize}

The user agent should not send any cookies in this context.  The user
agent may discard any cookie it receives in this context that the user
has not, through some user agent mechanism, deemed acceptable.
\end{quotation}

\subsubsection{Additive \textit{vs.} Independent Headers}
Dave Morris once again raised objections to the
additive solution to the compatibility problem in RFC 2109,
and Foteos Macrides joined him.
Macrides had actually implemented both RFC 2109 and the subsequent
I-Ds in the Lynx text-only browser,
and he felt the additive solution was highly error-prone,
both on the client side (matching the components of the respective
headers) and in applications (sending the corresponding pieces
correctly).

Recall that
the impetus for the additive approach was to
avoid sending the cookie value twice.
Two events prompted us to consider dropping the additive approach
and returning to the originally proposed and arguably
simpler-to-implement independent header approach:
\begin{enumerate}
\item Dave Morris described how a server would only need to send
both \verb,Set-Cookie, and \verb,Set-Cookie2, headers
the first time it receives a request from a client.
On subsequent requests, a client will send a
\verb,Cookie, header, and its content will reveal
whether the client understands v0 or v1 cookies.
The server can then send one response header, either
\verb,Set-Cookie, or \verb,Set-Cookie2,.
\item
Since Yaron Goland was the
lone voice arguing for the additive approach,
and assertions were made to \cite{http-wg} that neither
Microsoft nor Netscape would implement RFC 2109+,
there no longer
seemed to be a reason to pursue the unpopular and fragile additive approach.
\end{enumerate}

It might seem strange to continue a standards effort for a
feature that two major vendors say they will not support.
While that is certainly an undesirable circumstance, it does not necessarily
derail the process.
An IETF standard may represent what is considered the best
technical approach, even if vendors disagree.
Major vendors' voices do not trump ``rough consensus.''
Moreover, the requirement for at least two interoperative
implementations to exist before a Proposed Standard can
advance in the process does not require that 
they come from major vendors.
The issue is whether the standard can be implemented
consistently, not whether it is popular.

At the end of July, 1997, we were again bumping
up against an I-D submission deadline before a meeting in Munich.
On July 29 I recommended dropping the additive
approach in favor of the independent
header approach.

The goal of RFC 2109+ was to advance cookie technology
from \cite{NS} to something better-defined, more standard,
and with better privacy safeguards for users.
But a technical issue could impede the transition to this technology, namely,
how could a server discover that a user's browser supported
the newer technology?
As described above, a server could successfully learn that
a browser understood v1 cookies if it responded to a
request that contained no \verb,Cookie, header:
It would send both \verb,Set-Cookie, and \verb,Set-Cookie2, headers.
The browser's next request would reveal which of those
it understood.
But suppose the user upgraded her browser to one that understood
v1 cookies, but it retained an old cookie repository.
The cookies sent in new requests would still be v0 cookies,
and the server would not realize that the browser could
handle v1 cookies.

We quickly converged on a solution.
A browser sends
\verb,Cookie2: $Version=1, when it sends v0 cookies but it
understands v1 cookies.
\begin{quotation}
The Cookie2 header advises the server that the user agent understands
new-style cookies.  If the server understands new-style cookies, as
well, it should continue the stateful session by sending a Set-Cookie2
response header, rather than Set-Cookie.  A server that does not
understand new-style cookies will simply ignore the Cookie2 request
header.
\end{quotation}

On July 29 I announced an unofficial (not submitted) I-D for
inspection by the working group that contained the wording
that returned to the ``independent headers'' approach, along with minor wording
improvements regarding domain and host names.
My goal was to submit an acceptable I-D by the pre-meeting cutoff
the next day.
Indeed, after some minor comments (to add \verb,Cookie2,), I did
submit a new I-D, which was announced August 5, as
\verb,state-man-mec-03,.

\subsection{Munich IETF Meeting: August, 1997, and After}
I did not attend the Munich meeting, although both cookies
and ``certified cookies'' were on the agenda.
However, Judson Valeski
reported that the consensus
(actually, straw poll)
of people attending was to remove \verb,Comment, and \verb,CommentURL,
from the specification.
However, Larry Masinter pointed out that ``most
of the concerned parties weren't there''~\cite{http-wg}.
He went on to say
\begin{quotation}
It is my WG-chair opinion that progress on the protocol itself
has been held hostage to the current language in the protocol
description dealing with the privacy issue, and that one
way to make progress might be to split the document but not
the specification. (This would allow the privacy considerations
section to be revised even if the protocol specification
was not.)
\end{quotation}

\subsection{Splitting the Specification}
No further public discussion of cookies occurred for two months.
On October 10, 1997, I posted a message
to \cite{http-wg}\footnote{
``making progress on cookies''
}
to invite discussion on the proposal to split
the specification into two parts.
One part would describe the purely technical ``wire protocol.''
The second would address the privacy and ``unverifiable transaction''
pieces of the specification.
\begin{quotation}
To quote one of the Appl. Area Directors:  ``The point of serializing
these efforts is to focus the working group's discussion.''
\end{quotation}

After some private discussions with Keith Moore and Larry
Masinter about whether this approach would yield any progress,
I agreed to split the documents.
The plan was to reach consensus on the
first part before visiting the second,
and the first part would then be closed to discussion.
The resulting wire-protocol-only draft,
\verb,state-man-mec-04,, was announced
on October 25 and carried a note explaining that the
privacy provisions had been temporarily removed.
Over two weeks elapsed with no discussion of \verb,state-man-mec-04,
whatsoever.
Accordingly, Larry Masinter issued a working group Last Call
on November 11.
A few issues were raised on \cite{http-wg} of an essentially
editorial nature, and the comments were folded into another
I-D, \verb,state-man-mec-05,,
which anticipated yet another IETF meeting in early
December.

\subsection{Washington IETF Meeting: December, 1997, and after}
The HTTP Working Group sessions at the IETF meeting were
mostly concerned with finalizing the HTTP specification,
as this was likely to be the final meeting of the WG.
I gave a brief presentation on the current state
of the cookie specification.
Later, a small group of interested people
met informally to discuss the specification.
We agreed there was one technical issue remaining, and one political issue.
The political issue was the ever-popular ``unverifiable transactions.''

The technical issue was the domain-matching rule,
with two sub-problems:
\begin{enumerate}
\item how to restrict the set of servers to which a cookie
can be returned; and
\item how to support ``flat namespaces.''
For example, if an intranet had two hosts,
\verb,foo, and \verb,bar,, and if there were no
associated domain with those names, then
\verb,foo, and \verb,bar, would be unable to share cookies.
\end{enumerate}
On December 15, I posted a message to \cite{http-wg}~\footnote{
``the state of State''
}
to summarize where I thought things stood.
The first order of business was to resolve the intranet
cookie sharing issue.
On December 31 I started a thread in \cite{http-state}~\footnote{
``Step 1: domain matching rules''
}
to discuss it.
We considered Internet Explorer's ``Zones'' feature as
a model for how to share cookies
in an intranet, but we couldn't agree on how
to describe Zones in a technology-neutral way,
and Zones seemed to be addressing a much bigger issue anyway.
On January 2, Scott Lawrence
mentioned that another
working group was considering the
\verb,.local, domain as the implicit domain for
intranets.
We quickly converged on the idea of using
\verb,Domain=.local, to allow a server to
share cookies with all other servers in an intranet.
However, this solution did not provide a means to
restrict a cookie to some, but not all such servers.

Solving the broader domain-matching rules proved difficult
(as it had before).
On January 6, I summarized the dilemma~\cite{http-state}:
\begin{quotation}
So the challenge is to specify domain-matching rules that strike a
proper balance between simplicity and functionality (where functionality
includes allowing desirable outcomes and avoiding undesirable ones, such
as excessive cookie sharing).
 The rules must work correctly for .local
domain names.
\end{quotation}

On the one hand, it's easy to justify that
all servers belonging to one company in the domain \verb,example.com,
should be able to share cookies if they so desire.
In particular, these servers might all wish to
share cookies:
\begin{verbatim}
	example.com
	product.example.com
	v0.product.example.com
	v1.product.example.com
\end{verbatim}

On the other hand, imagine that a company \verb,mall.com,
hosts a ``shopping mall,''
so there are domain names \verb,shop1.mall.com,, \verb,shop2.mall.com,,
\textit{etc}.
Here it's obviously undesirable for the individual shops to be
able to see each other's cookies.

The underlying problem is that we're trying to infer the
bounds of administrative control based on domain names,
and that approach is inherently flawed.
The domain name system has
no externally imposed consistent structure.
Even \cite{NS}'s two-dot/three-dot rule is fragile
(and with the addition in 2000 of more top-level domains,
it would need to be extended).

Ultimately we agreed we could not resolve the intranet problem
beyond allowing a choice between ``share with all'' or ``share with none.''

\subsection{Working Group Last Call: March, 1998}
\verb,state-man-mec-06, drew no discussion.
Accordingly, on February 4, 1998, I attempted to
close off discussion of the protocol-only portion of the I-D.
There were a few further minor editorial comments
that led to \verb,state-man-mec-07,, which appeared on February 11.
Having completed the first part of the separate-drafts strategy,
I restored the privacy provisions to the specification,
which became available on February 18 as \verb,state-man-mec-08,.
Subsequent discussion was to focus solely on the privacy issues.

To my surprise, there were no further comments on the cookie
specification, despite the fact that I deliberately ``trolled''
for some.
Therefore I asked Larry Masinter to make a working group Last Call
for \verb,state-man-mec-08, to move forward on the standards track.
He asked to delay that step for a few days so he could simultaneously
issue a Last Call for two other working group items, one of which
was the HTTP/1.1 specification, and for all to advance to \emph{Draft}
Standard instead.
That Last Call was issued on March 13, 1998.

Advancing the cookie specification to Draft Standard would require
evidence of at least two independently written interoperating
implementations.
Accordingly I made private inquiries about people's implementations.
I learned that there were a few RFC 2109 client implementations
and one (nearly two) \verb,state-man-mec-08, client implementations.
However, no one volunteered that they had a server implementation
for either specification.
It therefore seemed appropriate procedurally to
deem it a new \emph{Proposed} Standard.
The IESG issued an IETF Last Call to that effect.

\subsection{Limbo:  April, 1998 to April, 2000}

The cookie specification then entered a nearly two-year limbo state.
Except for minor editorial changes, which led to new drafts
(through \verb,state-man-mec-12,), little of substance happened.
The mailing lists (\cite{http-state} and \cite{http-wg}) carried
virtually no discussion.
IANA requested a References section.
IESG requested that the typography for may/must/\ldots~language
use the more traditional IETF MAY/MUST/\ldots.

In June, 1998, I learned that the IESG was holding up the
progress of the specification to Proposed Standard.
In the year and a half since RFC 2109 had appeared, the
IESG's makeup had changed, and a newer member expressed
unhappiness that \verb,Comment, and \verb,CommentURL,
were not \emph{mandatory}, given the specification's
claimed devotion to privacy.
Ironically, the new IESG felt so strongly about the privacy
issues that, even though the wording in \verb,state-man-mec-08,
was at least as strict as in RFC 2109, they felt the need for
someone to write an ``applicability statement,''
because the working group and the IESG
could not agree on wording regarding privacy.
It would have words to the effect that
the cookie specification is approved as a Proposed Standard
subject to the condition that implementations \emph{also}
follow the applicability statement.
Keith Moore, Applications Area Director, finally wrote a draft
applicability statement,
\verb,draft-iesg-http-cookies-00.txt,,
in November.
Despite the appearance of the applicability statement,
the IESG did not act upon \verb,state-man-mec-10, 
through three IETF meetings and well past the draft's January, 1999,
nominal expiration.
Finally, on June 23, 1999, the IESG issued simultaneous IETF Last Calls
for the cookie specification and applicability statement.
Once again there was some perception, because of the long
delay, that these documents had sprung from nowhere.

The resulting discussion on \cite{http-wg}
included some minor technical comments and
some broad assertions.
Among the latter was an extended statement
that there was insufficient consensus
on the specification, that it was too controversial,
and that it should either be allowed to die silently or
should be deemed \emph{Experimental}.
Larry Masinter, as Working Group chair, held that process
had been followed, although he agreed that the consensus
was, indeed, ``rough.''

Over the next few weeks, a discussion developed on \cite{http-state}
about the technical issues, culminating, on August 17, in a new I-D,
\verb,state-man-mec-11,.
In response to worthwhile requests for clarifications from
an implementor, I made further revisions to the specifications,
and these appeared in
\verb,state-man-mec-12, on August 31.
Meanwhile, there had been comments on the applicability
statement as well, and it needed to be revised.

\subsection{Finale:  April, 2000 to October, 2000}
I began to ``ping'' the Area Directors periodically
about whether anything was happening.
Finally the IESG issued a Last Call for both
\verb,state-man-mec-12, and the applicability statement on April 28, 2000.
Their having received no further comments,
the IESG approved their advancement on August 7, 2000.
RFC 2965, \textit{HTTP State Management Mechanism}~\cite{rfc2965},
was announced on October 7.
RFC 2964, BCP (Best Current Practice) 44, \textit{Use of HTTP State Management},
was announced on October 12~\cite{rfc2964}.

\ifthenelse{\isDRAFT > 0}{\smallskip \appendixtexxSCCSID}{}
\else
\fi

\ifthenelse{\doTOC < 0} { \newpage \tableofcontents }{ }

\end{document}

